\documentclass[aps,prc,floatfix,preprint]{revtex4}

\preprint{JLAB-THY-11-1407}

\usepackage{amsmath}
\usepackage{graphicx}

\begin{document}

\title{Next-to-Leading Order Analysis of Target Mass Corrections \\
	to Structure Functions and Asymmetries}
\author{L. T. Brady$^{1,2}$, A. Accardi$^{1,3}$, T. J. Hobbs$^{1,4}$, 
	W. Melnitchouk$^1$}
\affiliation{
	$^1$Jefferson Lab, Newport News, Virginia 23606		\\
	$^2$Harvey Mudd College, Claremont, California 91711	\\
	$^3$Hampton University, Hampton, Virginia 23668		\\
	$^4$Indiana University, Bloomington, Indiana 47405}


\begin{abstract}
We perform a comprehensive analysis of target mass corrections (TMCs)
to spin-averaged structure functions and asymmetries at next-to-leading
order.  Several different prescriptions for TMCs are considered,
including the operator product expansion, and various approximations
to it, collinear factorization, and $\xi$-scaling.  We assess the impact
of each of these on a number of observables, such as the neutron to
proton $F_2$ structure function ratio, and parity-violating electron
scattering asymmetries for protons and deuterons which are sensitive to
$\gamma Z$ interference effects.  The corrections from higher order
radiative and nuclear effects on the parity-violating deuteron asymmetry
are also quantified.
\end{abstract}

\maketitle

\section{Introduction}
\label{sec:intro}

Tremendous progress has been made in recent years in the quest
to reveal the structure of the nucleon at its deepest levels.
Traditionally deep-inelastic scattering (DIS) of leptons has been
the primary tool used to study nucleon structure at large values
of the four-momentum transfer squared, $Q^2$, where the nucleon's
quark and gluon (or parton) constituents can be cleanly resolved.
Here the theoretical tools are well developed, and the nucleon's
structure can be conveniently parametrized in terms of universal
longitudinal momentum distribution functions of individual quarks
and gluons.  Global analyses of DIS and other hard processes have
been highly successful in correlating data over a wide range of
kinematics, producing fits of parton distribution functions (PDFs) to
next-to-leading order (in the strong coupling parameter $\alpha_s$)
accuracy and beyond \cite{MSTW08, JR09, ABKM10, HERAPDF10, CT10,
CTEQ6X, CJ11, NNPDF11}.

While the perturbative domain of large $Q^2$ and small parton momentum
fraction $x$ has received considerable attention both experimentally
and theoretically, the region of large $x$ and low $Q^2$
($\sim 1 - 2$~GeV$^2$), where nonperturbative effects play a greater
role, has been relatively poorly explored.  This is perhaps not too
surprising given the difficulty in reliably computing the various
corrections that are needed to describe data in this region.
Examples of nonperturbative effects that are relevant here include
target mass corrections (TMCs) associated with finite values of
$M^2/Q^2$, where $M$ is the nucleon mass, higher twist terms arising
from long-range nonperturbative multi-parton correlations, and
nuclear corrections in experiments involving deuterium or heavier
nuclei, which are important at large $x$ for any $Q^2$.

The large-$x$ region has been particularly difficult to access
experimentally, especially in high-energy colliders, due to the
rapidly falling cross sections as $x \to 1$.  The most extensive
data set available that covers this region has been from experiments
at SLAC \cite{SLAC}.  More recently, progress on this front has been
made with DIS structure function measurements at Jefferson Lab,
utilizing the high luminosities and duty factors available with the
CEBAF accelerator.  Indeed, an impressive body of very high-precision
data has now been accumulated over the last decade on various structure
functions, including accurate longitudinal--transverse separations
needed for model-independent determinations of the $F_2$ and $F_L$
structure functions of protons and nuclei \cite{Decade}.
Future plans at the energy upgraded Jefferson Lab involve extending
the DIS measurements to even larger $x$ ($x \sim 0.85$) with planned
experiments \cite{BONUS12, MARATHON, SOLID} to measure the ratio of
$d$ to $u$ quark distributions, as well as search for effects such as
charge symmetry violation in PDFs and tests of the Standard Model in
parity-violating DIS asymmetries.

The new data have the potential to provide strong constraints on PDFs
at large $x$, where currently uncertainties remain significant.
Several recent analyses \cite{CTEQ6X,CJ11} have in fact attempted to
utilize data at low values of $Q^2$ and invariant final state hadron
masses $W^2 = M^2 + Q^2 (1-x)/x$.  Stable fits of leading twist PDFs
could be obtained \cite{CTEQ6X} for $W$ down to $\sim 3$~GeV and
$Q^2 \gtrsim 1.7$~GeV$^2$, as long as TMC and higher twist corrections
were accounted for.
Aside from its intrinsic value, better knowledge of PDFs at large $x$
may also be important for searches of new physics signals in collider
experiments such as at the Tevatron or the LHC at large rapidities or
for heavy mass particles \cite{Brady11}, as well as at more central
rapidities where uncertainties in large-$x$ PDFs at low $Q^2$ can,
through $Q^2$ evolution, affect cross sections at small $x$ and large
$Q^2$ \cite{Kuhlmann00}.
%

The increased kinematic reach of the future high-precision DIS
measurements calls for a careful evaluation of the relevant
nonperturbative corrections in order to unambiguously extract
information on leading twist PDFs or new physics signals.
The effects that are most amenable to direct computation, in principle,
are the TMCs.  As discussed by Nachtmann \cite{Nachtmann73}, these
effects are in fact associated with leading twist operators (hence
contain no additional information on the nonperturbative parton
correlations), even though they give rise to
$Q^2/\nu^2 = 4 x^2 M^2/Q^2$ corrections, where $\nu = Q^2/2Mx$
is the energy transfer.  Nachtmann further showed that one could
generalize the standard operator product expansion (OPE) of
structure function moments to finite $Q^2$ such that only operators
of a specific twist would appear at a given order in $1/Q^2$.
The resulting target mass corrected structure functions can then
be derived through an inverse Mellin transformation, as shown by
Georgi and Politzer \cite{GP76} (for a review of TMCs in the OPE
approach see Ref.~\cite{Schienbein08}).

Later an alternative formulation in terms of collinear factorization
(CF) was used by Ellis, Furmanski and Petronzio \cite{EFP83} to derive
TMCs including the effects of off-shell partons and parton transverse
motion.  While the OPE and CF formulations yield identical results
for leading twist PDFs, they differ in the details of how the target
mass corrections are manifested at finite $Q^2$.  Other versions of
TMCs were subsequently derived \cite{AOT94,KR02,AQ08} within the CF
formalism using various assumptions about the intrinsic properties
of partons and higher twist contributions, leading to rather large
differences in some cases \cite{AQ08}.  Some of the phenomenological
implications of the different TMC prescriptions were discussed in
Refs.~\cite{KR02,AQ08}, including differences between leading
order and next-to-leading order (NLO) results; however, the effects
on observables have not been systematically investigated.  We do so
in this paper.

In Sec.~\ref{sec:TMC} we summarize the main results for TMCs in the
OPE and various CF formulations for the $F_1$, $F_2$, $F_3$ and $F_L$
structure functions at NLO, and illustrate the differences numerically.
Implications for various observables are discussed in Sec.~\ref{sec:obs},
including the ratio of neutron to proton $F_2$ structure functions,
which constrain the $d/u$ PDF ratio at large $x$, longitudinal
to transverse cross section ratios $R$, and parity-violating (PV)
DIS asymmetries on the proton and deuteron which are sensitive to
$\gamma Z$ interference structure functions.  We also quantify the
effects of perturbative NLO corrections on the $R^{\gamma Z}$ ratio
for the $\gamma Z$ interference, about which nothing is known
empirically, and of nuclear effects on the deuteron PV asymmetries.
Some finite-$Q^2$ effects on PV asymmetries were investigated
previously in Ref.~\cite{Hobbs08, Hobbs11}, and higher-twists in
deuteron PV asymmetries in Refs.~\cite{Bj78, Fajfer84, Castorina85,
Mantry10, Belitsky11}.  Finally, in Sec.~\ref{sec:conc} we draw some
conclusions and outline possible extensions of this work.

\section{Target Mass Corrections}
\label{sec:TMC}

In this section we review the kinematic corrections to structure
functions arising from scattering at finite values of $Q^2/\nu^2$.
We consider several frameworks for the TMCs, including the
conventional one based on the operator product expansion,
and various approximations to it, as well as a number of prescriptions
using collinear factorization at leading and next-to-leading order
in $\alpha_s$.
The structure functions for the scattering of an unpolarized lepton
from an unpolarized nucleon are defined in terms of the nucleon
hadron tensor \cite{PDG10},
\begin{subequations}
\label{eq:Wmunu}
\begin{eqnarray}
W_{\mu\nu}
&=& {1 \over 4 \pi} \int d^4 z\,e^{i q \cdot z}
    \langle p \left|
	\left[ J_\mu^\dagger(z), J_\nu(0) \right]
    \right| p \rangle
\label{eq:Wmunu_def}					\\
&=& \left( -g_{\mu\nu} + \frac{q_\mu q_\nu}{q^2} \right) F_1(x,Q^2)\
 +\ \left( p_\mu - \frac{p\cdot q}{q^2} q_\mu \right)
    \left( p_\nu - \frac{p\cdot q}{q^2} q_\nu \right)
    {F_2(x,Q^2) \over p \cdot q}			\nonumber\\
& &
 -\ i\epsilon_{\mu\nu\alpha\beta} q^{\alpha} p^{\beta}
    \frac{F_3(x,Q^2)}{2 p\cdot q}\, ,
\end{eqnarray}
\end{subequations}
where $J_\mu$ is the electromagnetic or weak current operator
for a given virtual boson ($\gamma$, $Z$ or $W^\pm$).
Here $p$ and $q$ are the nucleon and exchanged boson four-momenta,
respectively, with $q^2 = -Q^2$.

The structure functions $F_{1,2}$ are related to the product of
two vector or two axial-vector currents, while $F_3$ arises from the
interference of vector and axial-vector currents.  The $F_1$ structure
function is proportional to the transverse virtual boson cross section,
and $F_2$ is given by a combination of transverse and longitudinal
cross sections.  It is convenient to also introduce the longitudinal
structure function,
\begin{equation}
\label{eq:FLdef}
F_L(x,Q^2) = \rho^2 F_2(x,Q^2) - 2x F_1(x,Q^2)\, ,
\end{equation}
where
\begin{equation}
\rho^2 = 1 + \frac{4 x^2 M^2}{Q^2}.
\end{equation}
In the following we will summarize target mass corrections for each
of these structure functions computed within the various approaches
outlined above.

\subsection{Operator product expansion}
\label{ssec:OPE}

Target mass corrections to structure functions were first systematically
considered by Georgi and Politzer \cite{GP76} in the framework of the
operator product expansion.  Here the twist-two quark bilinears in the
product of currents $J_\mu J_\nu$ in Eq.~(\ref{eq:Wmunu_def}) are
modified with the introduction of covariant derivatives,
$\bar{\psi}\gamma^{\mu} D^{\mu_1} \cdots D^{\mu_n} \psi$; since each
derivative $D^{\mu_i}$ increases both the dimension and spin of the
operator by one unit, the twist (dimension minus spin) remains
unchanged.  The expansion in terms of covariant derivatives yields
a series in $1/Q^2$ with coefficients given by moments of structure
functions.  The resulting target mass corrected structure functions are
then accessed through an inverse Mellin transformation, which gives
\cite{GP76}
\begin{subequations}
\label{eq:OPE}
\begin{eqnarray}
F_1^{\rm OPE}(x,Q^2)
&=& {1+\rho \over 2\rho} F_1^{(0)}(\xi,Q^2)\
 +\ {\rho^2-1 \over 4\rho^2}
    \left[ h_2(\xi,Q^2) + {\rho^2-1 \over 2 x \rho} g_2(\xi,Q^2)
    \right],						\\
\label{eq:OPE_F2}
F_2^{\rm OPE}(x,Q^2)
&=& {(1+\rho)^2 \over 4 \rho^3} F_2^{(0)}(\xi,Q^2)\
 +\ {3x(\rho^2-1) \over 2\rho^4}
    \left[ h_2(\xi,Q^2) + {\rho^2-1 \over 2 x \rho} g_2(\xi,Q^2)
    \right],                                            \\
F_L^{\rm OPE}(x,Q^2)
&=& {(1+\rho)^2\over 4 \rho} F_L^{(0)}(\xi,Q^2)
 +\ {x(\rho^2-1) \over \rho^2}
    \left[ h_2(\xi,Q^2) + {\rho^2-1 \over 2 x \rho} g_2(\xi,Q^2)
    \right],                                            \\
\label{eq:OPE_F3}
F_3^{\rm OPE}(x,Q^2)
&=& {(1+\rho) \over 2 \rho^2} F_3^{(0)}(\xi,Q^2)
 +\ {(\rho^2-1) \over 2\rho^3} h_3(\xi,Q^2),
\end{eqnarray}
\end{subequations}
where $F_i^{(0)}$ are the structure functions in the $M^2/Q^2 \to 0$
limit, evaluated at the modified scaling variable $\xi$
\cite{Nachtmann73,Greenberg71},
\begin{equation}
\xi = {2x \over 1 + \rho}\, ,
\end{equation}
which approaches $x$ as $M^2/Q^2 \to 0$.
The functions $h_2$, $g_2$ and $h_3$ are associated with higher order
terms in $M^2/Q^2$ and are given by \cite{GP76,Schienbein08}
\begin{subequations}
\label{eq:hg}
\begin{eqnarray}
h_2(\xi,Q^2)
&=& \int_\xi^1 du\, {F_2^{(0)}(u,Q^2) \over u^2}\, ,\\
g_2(\xi,Q^2)
&=& \int_\xi^1 du\, \int_u^1 dv {F_2^{(0)}(v,Q^2) \over v^2}\
 =\ \int_\xi^1 du\, (u-\xi) {F_2^{(0)}(u,Q^2) \over u^2}\, ,\\
h_3(\xi,Q^2)
&=& \int_\xi^1 du\, {F_3^{(0)}(u,Q^2) \over u}\, .
\end{eqnarray}
\end{subequations}
(Note that the function $g_2$ here should not be confused with the
spin-dependent $g_2$ structure function measured in polarized
lepton--nucleon scattering.)

The expressions in Eqs.~(\ref{eq:OPE}) are known to suffer from the
``threshold problem'', in which the target mass corrected (leading
twist) structure functions do not vanish as $x \to 1$, and are in
fact nonzero in the kinematically forbidden $x > 1$ region, where
for a proton target baryon number conservation would be violated.
This is clear from the ${\cal O}(1)$ terms in Eqs.~(\ref{eq:OPE})
in which the massless functions $F_i^{(0)}$ are evaluated at $\xi$.
Because at any finite $Q^2$ value one has
	$\xi < \xi_0 \equiv \xi(x=1) < 1$,
for any input function $F_i^{(0)}$ which is nonzero for $0 < x < 1$,
the target mass corrected function at $x=1$ will not vanish,
$F_i^{\rm OPE}(x=1,Q^2<\infty) > 0$.
A number of attempts have been made to ameliorate the threshold
problem \cite{Tung79,Steffens06} using various prescriptions and
{\it ans\"atze}, although none of these is unique and without
additional complications \cite{Schienbein08}.

Recently, Kulagin and Petti \cite{KP06} showed that by expanding the
target mass corrected structure functions to leading order in $1/Q^2$,
the resulting functions have the correct $x \to 1$ limits,
\begin{subequations}
\label{eq:KP1}
\begin{eqnarray}
F_1^{1/Q^2}(x,Q^2)
&=&
\frac{1}{4}
\left(5-\rho^2\right)F_1^{(0)}(x,Q^2)\
-\ \frac{1}{4} \left(\rho^2-1\right)
\left[ x F_1^{(0)\, \prime}(x,Q^2)
     -\ h_2(x,Q^2)
\right],						\\
\label{eq:KP1_F2}
F_2^{1/Q^2}(x,Q^2)
&=&
\left(2-\rho^2\right)F_2^{(0)}(x,Q^2)\
-\ \frac{1}{4} \left(\rho^2-1\right)
\left[ x F_2^{(0)\, \prime}(x,Q^2)
     -\ 6 x h_2(x,Q^2)
\right],						\\
F_L^{1/Q^2}(x,Q^2)
&=&
F_L^{(0)}(x,Q^2)\
-\ \frac{1}{4} \left(\rho^2-1\right)
\left[ x F_L^{(0)\, \prime}(x,Q^2)
     -\ 4 x h_2(x,Q^2)
\right],						\\
F_3^{1/Q^2}(x,Q^2)
&=&
\frac{1}{4}
\left(7-3\rho^2\right)F_3^{(0)}(x,Q^2)\
-\ \frac{1}{4} \left(\rho^2-1\right)
\left[ x F_3^{(0)\, \prime}(x,Q^2)
     -\ 2 h_3(x,Q^2)
\right].						\nonumber\\
& &
\end{eqnarray}
\end{subequations}
While avoiding the threshold problem, this prescription, however, raises
the question of whether the $1/Q^2$ approximation is sufficiently
accurate for structure functions near $x \approx 1$ at moderate $Q^2$.
To test the convergence of the $1/Q^2$ expansion at large $x$,
we further expand the OPE results (\ref{eq:OPE}) to include
${\cal O}(1/Q^4)$ corrections,
\begin{subequations}
\label{eq:KP2}
\begin{eqnarray}
F_1^{1/Q^4}(x,Q^2)
&=& F_1^{1/Q^2}(x,Q^2)\ +\ 
\left( \rho^2-1 \right)^2
\left[
    \frac{3}{16}  F_1^{(0)}(x,Q^2)\
 +\ \frac{1}{16x} F_2^{(0)}(x,Q^2)
\right.							\nonumber\\
&+&
\left.
    \frac{3x}{16} F_1^{(0)\, \prime}(x,Q^2)\
 +\ \frac{x^2}{32}\, F_1^{(0)\, \prime\prime}(x,Q^2)\
 -\ \frac{1}{4}\,  h_2(x,Q^2)\
 +\ \frac{1}{8x}\, g_2(x,Q^2)
\right]							\\
\label{eq:KP2_F2}
F_2^{1/Q^4}(x,Q^2)
&=& F_2^{1/Q^2}(x,Q^2)\ +\ 
\left( \rho^2-1 \right)^2
\left[
    \frac{23}{16} F_2^{(0)}(x,Q^2)\
 +\ \frac{3x}{8}  F_2^{(0)\, \prime}(x,Q^2)
\right.							\nonumber\\
&+&
\left.
    \frac{x^2}{32}\, F_2^{(0)\, \prime\prime}(x,Q^2)\
 -\ 3x\, h_2(x,Q^2)\
 +\ \frac{3}{4}\, g_2(x,Q^2)
\right],						\\
F_L^{1/Q^4}(x,Q^2)
&=& F_L^{1/Q^2}(x,Q^2)\ +\ 
\left( \rho^2-1 \right)^2
\left[
    \frac{3}{16} F_L^{(0)}(x,Q^2)\		
 +\ \frac{1}{4}  F_2^{(0)}(x,Q^2)
\right.							\nonumber\\
&+&
\left.
    \frac{x}{8}    F_L^{(0)\, \prime}(x,Q^2)\
 +\ \frac{x^2}{32} F_L^{(0)\, \prime\prime}(x,Q^2)\
 -\ x h_2(x,Q^2)\
 +\ \frac{1}{2} g_2(x,Q^2)
\right],						\\
F_3^{1/Q^4}(x,Q^2)
&=& F_3^{1/Q^2}(x,Q^2)\ +\ 
\left( \rho^2-1 \right)^2
\left[
    \frac{13}{16} F_3^{(0)}(x,Q^2)	
 +\ \frac{5x}{16} F_3^{(0)\, \prime}(x,Q^2)
\right.							\nonumber\\
&+&
\left.
    \frac{x^2}{32}\, F_3^{(0)\, \prime\prime}(x,Q^2)\
 -\ \frac{3}{4}\, h_3(x,Q^2)
\right],
\end{eqnarray}
\end{subequations}
where the first $(F_i^{(0)\, \prime})$ and second
$(F_2^{(0)\, \prime\prime})$ derivatives of the structure functions
are with respect to $x$.  In fact, one can show that for a structure
function that behaves at large $x$ as $(1-x)^n$, the target mass
corrected result will vanish in the $x \to 1$ limit up to order
$1/Q^{2n-2}$ in the expansion.  For $n \approx 3$, as is typical for
nucleon structure functions, the threshold problem will therefore
appear only at order $1/Q^6$.

\begin{figure}[t]
\rotatebox{-90}{\includegraphics[width=7cm]{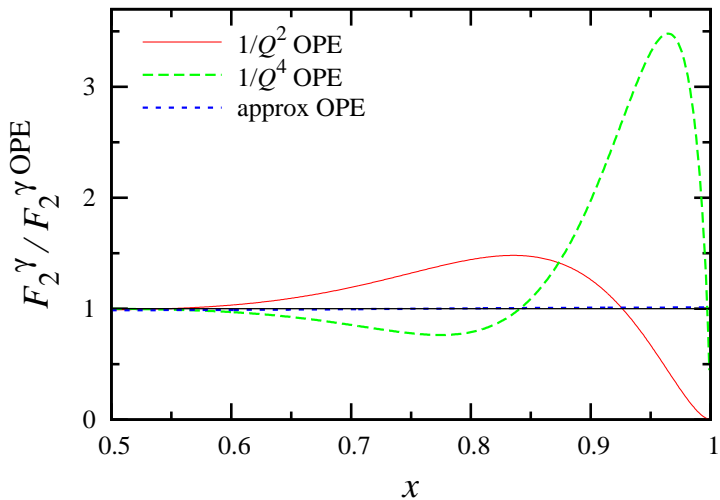}}
\caption{Ratio of the target mass corrected $F_2$ structure functions
	using the $1/Q^2$ (solid, red), $1/Q^4$ (long-dashed, green)
	and phenomenological (short-dashed, blue) OPE approximations
	compared with the exact OPE result, Eq.~(\ref{eq:OPE_F2}).
	Note that the phenomenological OPE approximation is almost
	indistinguishable from the exact OPE result, while the $1/Q^2$
	and $1/Q^4$ expansions deviate from this for $x \gtrsim 0.6$.}
\label{fig:OPE_comp}
\end{figure}

The accuracy of the $1/Q^2$ expansions is illustrated in
Fig.~\ref{fig:OPE_comp}, where in order to isolate the target mass
effect from the specific form of the structure function parametrization
we have taken for simplicity the form $F_2 \sim (1-x)^3$.
Both the $1/Q^2$ and $1/Q^4$ approximations are found to reproduce
the full OPE result very well up to $x \approx 0.6$, but significant
deviations are visible at larger $x$.  Furthermore, while there is
a modest improvement in the agreement with the exact result for
$0.6 \lesssim x \lesssim 0.8$ after inclusion of the $1/Q^4$ terms,
both expansions appear to break down for $x \gtrsim 0.8$.
The reliability of a low order $1/Q^2$ expansion is therefore
questionable at these $x$ values, and hence their efficacy in
removing the $x \to 1$ threshold problem.

Since the integrals in the functions $h_{2,3}$ and $g_2$ can be
time consuming to compute numerically, Schienbein {\it et al.}
\cite{Schienbein08} found phenomenological analytic forms which
approximate the target mass corrected $F_2$ and $F_3$ structure
functions in Eqs.~(\ref{eq:OPE_F2}) and (\ref{eq:OPE_F3}) by
\begin{subequations}
\label{eq:approxOPE}
\begin{eqnarray}
\label{eq:approxOPE_F2}
F_2^{\rm approx}(x,Q^2)&=&\frac{(1+\rho)^2}{2\rho^3}
\left(1+\frac{3(\rho^2-1)}{\rho(1+\rho)}(1-\xi)^2\right)
F_2^{(0)}(\xi,Q^2),\						\\
F_3^{\rm approx}(x,Q^2)&=&\frac{(1+\rho)}{2\rho^2}
\left(1-\frac{(\rho^2-1)}{2\rho(1+\rho)}(1-\xi)\ln{\xi}\right)
F_3^{(0)}(\xi,Q^2).
\end{eqnarray}
\end{subequations}
These turn out to be rather good approximations to the exact results,
as Fig.~\ref{fig:OPE_comp} illustrates for the $F_2$ case.  For all
values of $x$, the phenomenological approximation (\ref{eq:approxOPE_F2})
stays within 5\% of the full OPE result.

\subsection{Collinear factorization}
\label{ssec:CF}

An alternative approach to TMCs relies on the collinear factorization
(CF) formalism \cite{EFP83,AOT94,KR02,AQ08}, which makes use of the
factorization theorem to relate the hadronic tensor for lepton--hadron
scattering to that for scattering from a parton.  Here parton
distributions are formulated directly in momentum space, avoiding the
need to perform an inverse Mellin transform to obtain the PDF from its
moments.  An advantage of the CF formalism for TMCs is that it can be
extended to other hard scattering processes, such as semi-inclusive DIS
\cite{AHM09}, where an OPE is not available.

\subsubsection{Ellis, Furmanski and Petronzio}
\label{sssec:EFP}

The first study of TMCs within CF was made by Ellis, Furmanski, and
Petronzio (EFP) \cite{EFP83}, who analyzed the virtual photon-hadron
scattering amplitude using a Feynman diagram technique to expand the
hard scattering term about the collinear direction, incorporating
both parton off-shellness (or interactions) and parton transverse
momentum in twist-4 contributions \cite{Qiu90}.
Using the same notation as for the OPE TMCs above, the EFP results
for the target mass corrected structure functions are given by
\begin{subequations}
\label{eq:EFP}
\begin{eqnarray}
F_1^{\rm EFP}(x,Q^2)&=&
\frac{2}{1+\rho} F_1^{(0)}(\xi,Q^2)
+ \frac{(\rho^2-1)}{(1+\rho)^2} h_2(\xi,Q^2)\, ,	\\
F_2^{\rm EFP}(x,Q^2)&=&
\frac{1}{\rho^2} F_2^{(0)}(\xi,Q^2) + \frac{3\xi(\rho^2-1)}
{\rho^2 (1+\rho)} h_2(\xi,Q^2)\, ,			\\
F_L^{\rm EFP}(x,Q^2)&=&
F_L^{(0)}(\xi,Q^2)
+ \frac{2\xi(\rho^2-1)}{(1+\rho)} h_2(\xi,Q^2)\, ,	\\
F_3^{\rm EFP}(x,Q^2)&=&
\frac{1}{\rho} F_3^{(0)}(\xi,Q^2)+\frac{2(\rho^2-1)}
{\rho (1+\rho)^2} h_3(\xi,Q^2)\, ,
\end{eqnarray}
\end{subequations}
where again the $F_i^{(0)}$ refer to the uncorrected structure
functions, and $h_{2,3}$ are given in Eqs.~(\ref{eq:hg}).
(Note that the definition of the longitudinal structure function in
EFP differs from the usual definition (\ref{eq:FLdef}) by a factor
$x$, and the $F_2$ structure function is proportional to what EFP
call the ``transverse'' structure function, which in standard usage
is proportional to $F_1$.)
Because the massless functions $F_i^{(0)}$ are evaluated at $\xi$,
the target mass corrected structure functions will suffer from the
same threshold problem as in the OPE analysis in Eqs.~(\ref{eq:OPE}).
While the expressions in Eqs.~(\ref{eq:EFP}) were derived in
Ref.~\cite{EFP83} at leading order in $\alpha_s$, in this work
we will assume their validity also at NLO.

The prefactors for the leading terms proportional to $F_i^{(0)}$
in Eqs.~(\ref{eq:EFP}) are remarkably close to those for the leading
terms in the OPE expressions in Eqs.~(\ref{eq:OPE}).  To first order
in $1/Q^2$, the leading term prefactors for $F_1$ in both OPE and EFP
reduce to $(1-x^2 M^2/Q^2)$.  Similarly, the $F_2$ prefactors both
reduce to $(1-4 x^2 M^2/Q^2)$, while those for $F_L$ reduce to 1.
For the $F_3$ structure function, however, the ${\cal O}(1/Q^2)$
prefactor is $(1-3 x^2 M^2/Q^2)$ for OPE, whereas for the EFP CF
result it is $(1-2 x^2 M^2/Q^2)$.

At leading order in the massless limit the longitudinal structure
function vanishes identically.  At NLO, however, it receives
contributions from both quark and gluon PDFs convoluted with the
respective hard coefficient functions.  For electromagnetic
scattering, for example, one has \cite{Altarelli78,Bardeen78}
\begin{eqnarray}
F_L^{\gamma (0)}(x,Q^2)
&=& {\alpha_s(Q^2) \over \pi}
    \int_x^1 {dy \over y} \left( {x \over y} \right)^2
    \left\{ {4 \over 3} F_2^{\gamma (0), \rm LO}(y,Q^2)
	  + c^\gamma (y-x) g(y,Q^2)
    \right\},
\end{eqnarray}
where $c^\gamma = 2\sum_q e_q^2$, and $F_2^{\gamma (0), \rm LO}$
is given by the leading order expression for $F_2^{\gamma (0)}$.
Similar expressions hold also for the longitudinal structure functions
associated with other electroweak currents.  In our numerical
calculations discussed below we will always compute $F_L$ at NLO.

It is important also to note that Eqs.~(\ref{eq:EFP}) have been
derived considering Feynman diagrams with 2 or 4 legs attached to
the hadronic correlator (see Figs.~2 and 3 of Ref.~\cite{EFP83}),
which for $M=0$ give rise to twist-2 and twist-4 contributions to
the structure functions, respectively.  For $M \neq 0$, however,
the quark and gluon equations of motion allow one to extract a
twist-2 contribution from the 4-leg diagrams, which when added
to the twist-2 target mass correction yields the full result in
Eqs.~(\ref{eq:EFP}).  It is an interesting question whether by
resumming the twist-2 parts of $n$-leg diagrams one would be able
to recover the TMC expressions (\ref{eq:EFP}).

\subsubsection{Accardi and Qiu}
\label{sssec:AQ}

In both the EFP and OPE treatments of TMCs, the resulting structure
functions are nonzero for $x>1$.  The analysis of Accardi and Qiu (AQ)
\cite{AQ08} traced this problem to baryon number nonconservation in
the handbag diagram for $M \neq 0$.  Working with 2-leg diagrams only,
in contrast to EFP who also consider 4-leg diagrams up to twist-4, the
AQ target mass corrected structure functions are given by \cite{AQ08}
\begin{subequations}
\label{eq:AQ}
\begin{eqnarray}
F_1^{\rm AQ}(x,Q^2)
&=& \widetilde{F}_1^{(0)}(\xi,Q^2),			\\
F_2^{\rm AQ}(x,Q^2)
&=& \frac{1+\rho}{2\rho^2} \widetilde{F}_2^{(0)}(\xi,Q^2),	\\
F_L^{\rm AQ}(x,Q^2)
&=& \frac{1+\rho}{2} \widetilde{F}_L^{(0)}(\xi,Q^2),	\\
F_3^{\rm AQ}(x,Q^2)
&=& \frac{1}{\rho} \widetilde{F}_3^{(0)}(\xi,Q^2).
\end{eqnarray}
\end{subequations}
Here the functions $\widetilde{F}_i^{(0)}$ are defined as
\begin{align}
\label{eq:AQ2}
\widetilde{F}_i^{(0)}(\xi,Q^2)
= \sum_f \int_\xi^{\xi/x} \!\frac{dz}{z}\,
  C_i^f\left({\xi\over z},Q^2\right) \varphi_f(z,Q^2),
\end{align}
where $C_i^f$ are the perturbatively calculable hard coefficient
functions for a given parton flavor $f$, including parton charge
factors, $\varphi_f$ are the parton densities of the nucleon,
and the sum is taken over all active flavors.
The upper limit in Eq.~(\ref{eq:AQ2}) ensures that the target mass
corrected structure functions vanish for $x > 1$, as required by
kinematics, although jet mass corrections need to be introduced in
order to render the target mass corrected functions zero at $x=1$
\cite{AQ08}.
It remains an interesting exercise to apply the same prescription
to twist-4 diagrams as in Ref.~\cite{EFP83} in order to establish a
more direct correspondence between the AQ and EFP approaches.
Of course, for $M^2/Q^2 \to 0$ the upper limit of integration in
Eq.~(\ref{eq:AQ2}) is 1, and both approaches recover the standard
factorization theorem for structure functions~\cite{CSS88}.

\subsubsection{$\xi$-scaling}
\label{sssec:xi}

When the upper limit of integration in Eq.~(\ref{eq:AQ2}) is taken to
be 1, the AQ structure functions reduce to the simple $\xi$-scaling
($\xi$-S) form introduced by Aivazis {\it et al.} \cite{AOT94} and
used by Kretzer and Reno \cite{KR02}.  The target mass corrected
structure functions in this case are simply given by
\begin{subequations}
\label{eq:xiS}
\begin{eqnarray}
F_1^{\rm \xi\mbox{-}S}(x,Q^2)
&=& F_1^{(0)}(\xi,Q^2)\, ,				\\
F_2^{\rm \xi\mbox{-} S}(x,Q^2)
&=& \frac{1+\rho}{2 \rho^2}\ F_2^{(0)}(\xi,Q^2)\, ,	\\
F_L^{\rm \xi\mbox{-} S}(x,Q^2)
&=& \frac{1+\rho}{2}F_L^{(0)}(\xi,Q^2)\, ,			\\
F_3^{\rm \xi\mbox{-} S}(x,Q^2)
&=& \frac{1}{\rho} F_3^{(0)}(\xi,Q^2)\, .
\end{eqnarray}
\end{subequations}%
Note that the form of the target mass corrected functions in
Eqs.~(\ref{eq:xiS}) closely resembles that in Eqs.~(\ref{eq:AQ}),
with the two forms equivalent at leading order.
At this order the structure functions satisfy a modified Callan-Gross
relation \cite{AQ08},
\begin{align}
\rho^2\, F_2^{\rm \xi\mbox{-} S}(x,Q^2)
= 2x F_1^{\rm \xi\mbox{-} S}(x,Q^2)\, .
\end{align}
The leading order $\xi$-scaling structure functions are also related
to the leading, ${\cal O}(1)$ terms of the OPE expressions in
Eqs.~(\ref{eq:OPE}),
\begin{equation}
F_i^{\rm OPE\, (leading)}(\xi,Q^2)
= \frac{1+\rho}{2\rho}\ F_i^{\rm \xi\mbox{-} S}(\xi,Q^2)\, ,
\end{equation}
where the prefactor, to order $1/Q^2$, is given by $(1-x^2 M^2/Q^2)$.
In fact, the $\xi$-scaling formulas (\ref{eq:xiS}) would coincide
with the EFP results in Sec.~\ref{sssec:EFP} in the absence of 4-leg 
Feynman diagrams \cite{EFP83}.

\subsection{TMC comparisons}
\label{sssec:comp}

\begin{figure}[thb]
\rotatebox{-90}{\includegraphics[width=5.7cm]{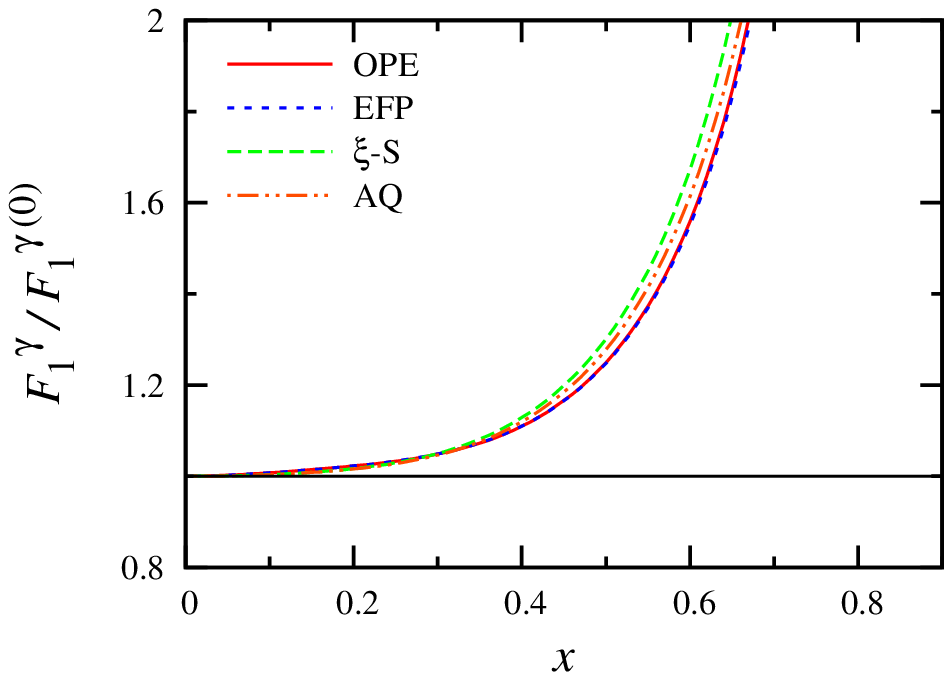}}
\rotatebox{-90}{\includegraphics[width=5.7cm]{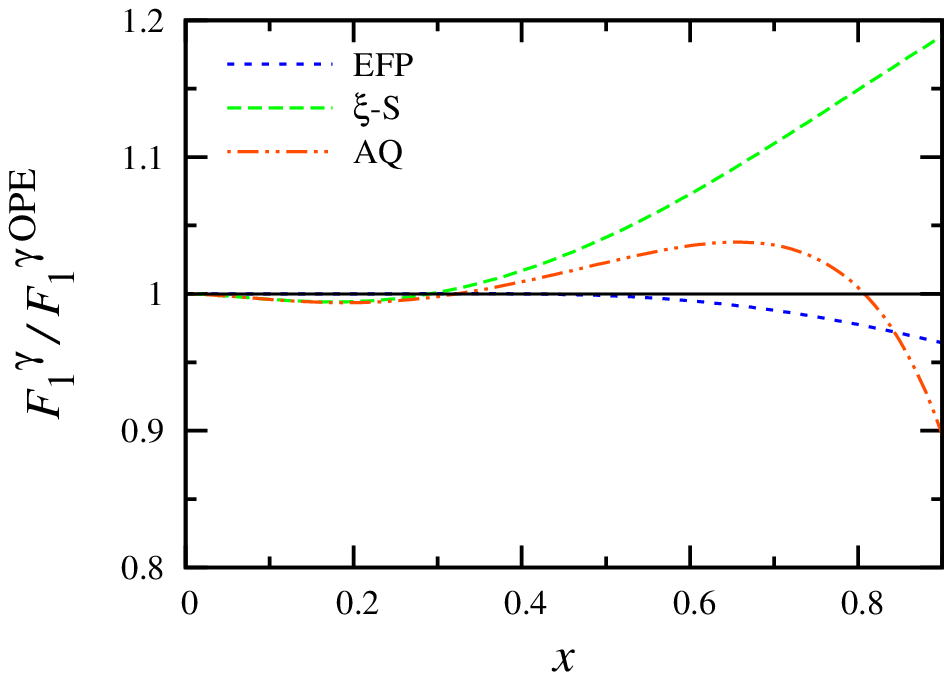}}
\caption{Ratio of target mass corrected to uncorrected {\bf (left)}
	or to OPE {\bf (right)} $F_1^\gamma$ proton structure
	functions at $Q^2 = 2$~GeV$^2$ for the
	OPE (solid, red),
	EFP (short-dashed, blue),
	$\xi\mbox{-S}$ (long-dashed, green), and
	AQ (dot-dashed, orange) TMC prescriptions.
	Note that the OPE and EFP results are almost 
	indistinguishable for $x \lesssim 0.6$.}
\label{fig:F1}
\end{figure}

\begin{figure}[hb]
\rotatebox{-90}{\includegraphics[width=5.7cm]{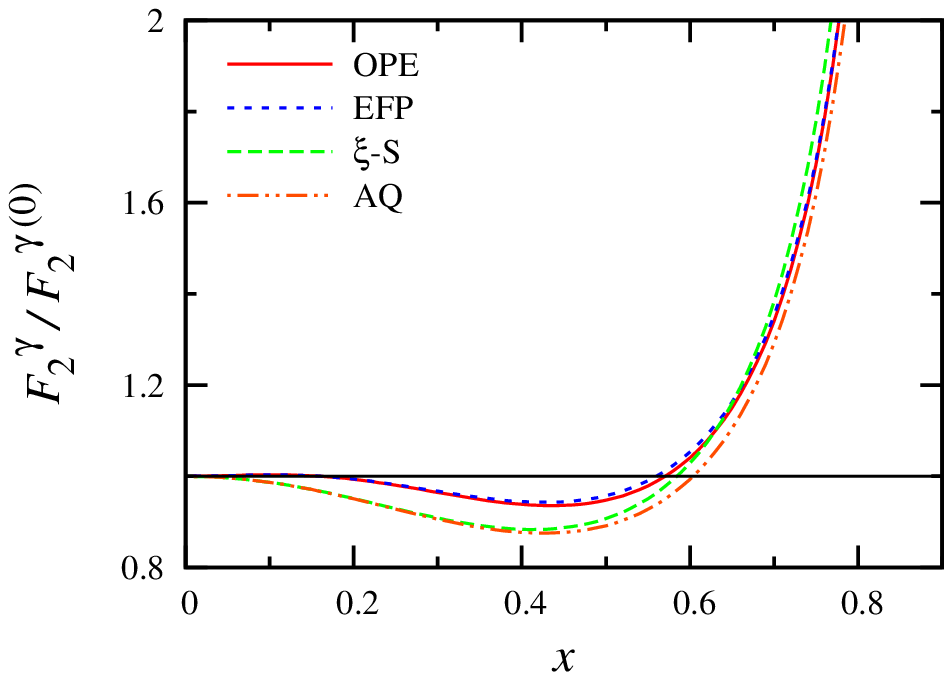}}
\rotatebox{-90}{\includegraphics[width=5.7cm]{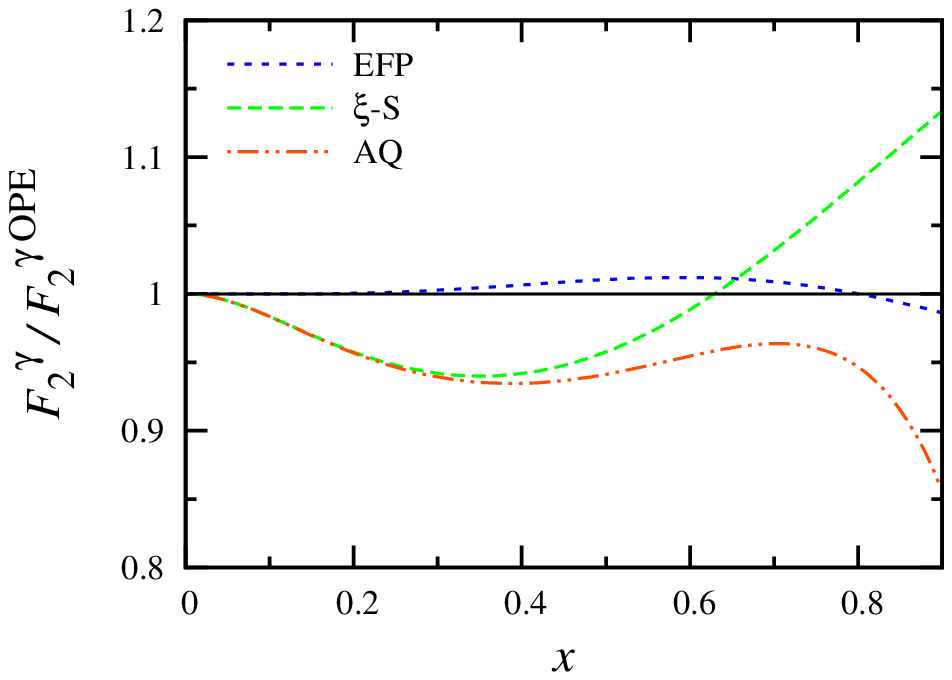}}
\caption{As in Fig.~\ref{fig:F1}, but for the $F_2^\gamma$ proton
	structure function.}
\label{fig:F2}
\end{figure}

\begin{figure}[t]
\rotatebox{-90}{\includegraphics[width=5.7cm]{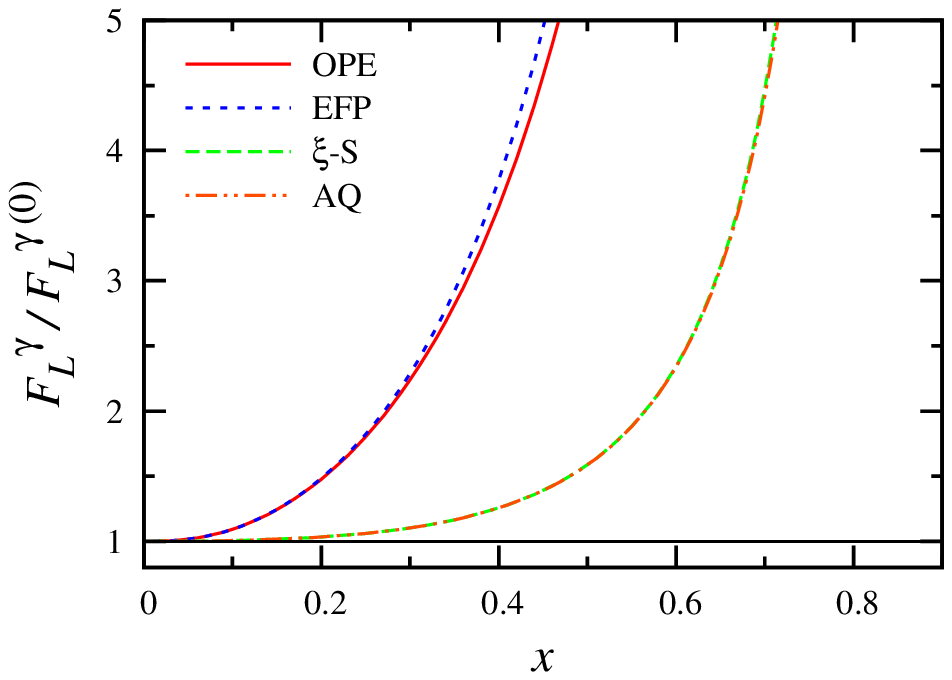}}
\rotatebox{-90}{\includegraphics[width=5.7cm]{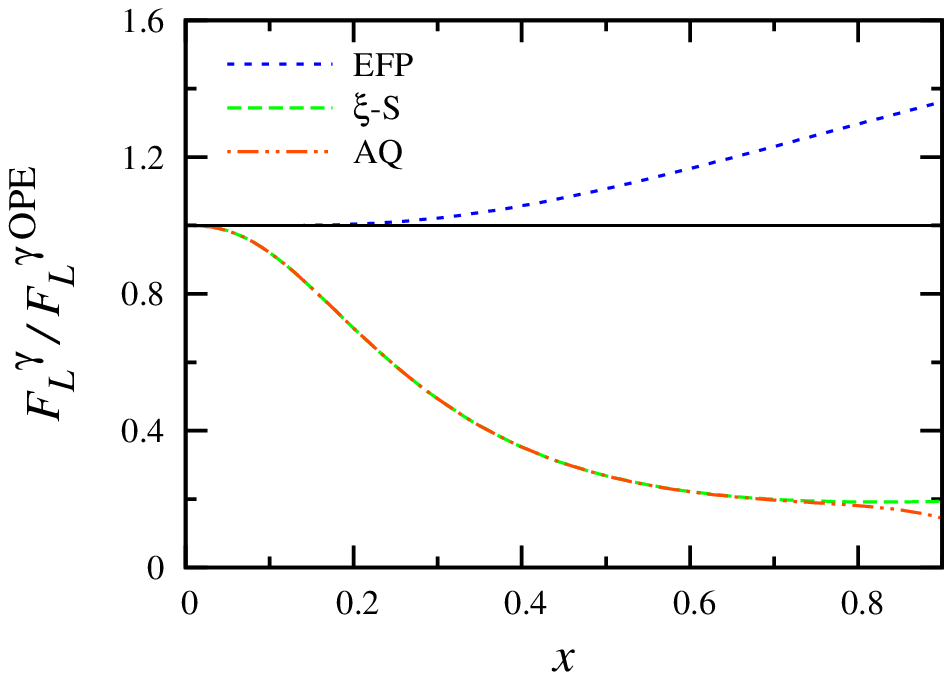}}
\caption{As in Fig.~\ref{fig:F1}, but for the $F_L^\gamma$ proton
	structure function.}
\label{fig:FL} 
\end{figure}

\begin{figure}[h]
\rotatebox{-90}{\includegraphics[width=5.7cm]{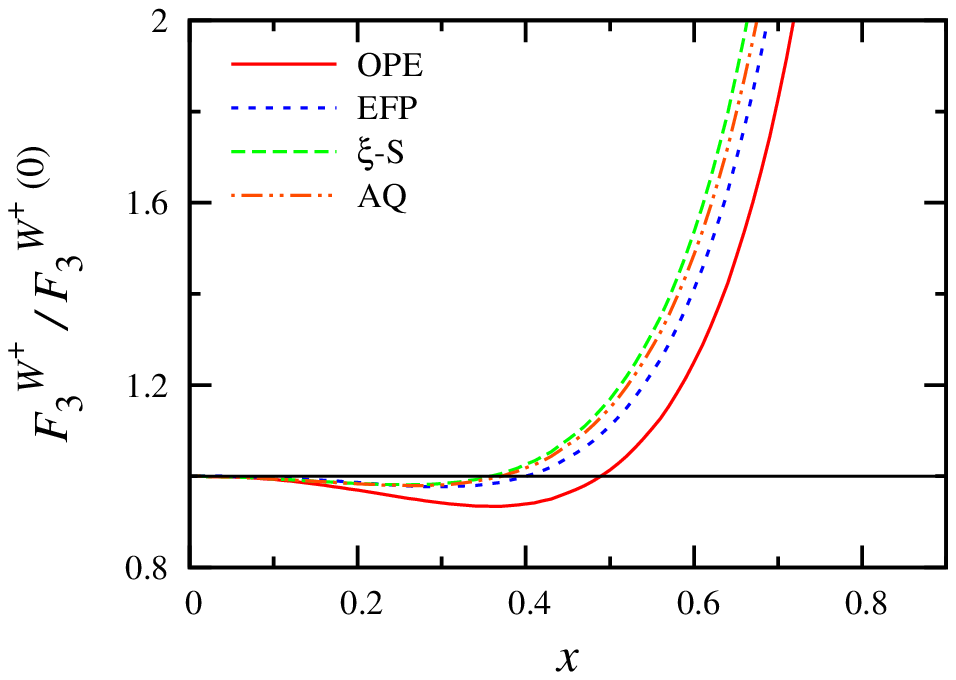}}
\rotatebox{-90}{\includegraphics[width=5.7cm]{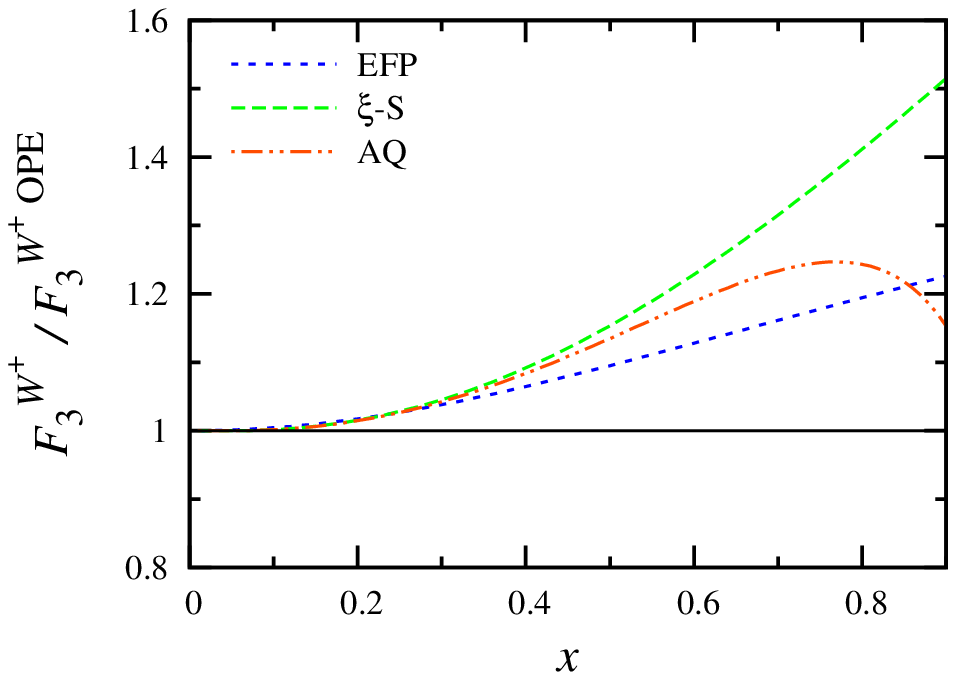}}
\caption{As in Fig.~\ref{fig:F1}, but for the $F_3^{W^+}$ proton
        structure function.}
\label{fig:F3}
\end{figure}

The effects of the different TMC prescriptions on structure
functions are illustrated in Figs.~\ref{fig:F1} -- \ref{fig:F3}
for the $F_1^\gamma$, $F_2^\gamma$, $F_L^\gamma$ and $F_3^{W+}$
structure functions of the proton, respectively.
(The results for structure functions associated with other
boson exchanges, such as $W^-$, $Z$, or $\gamma Z$ interference,
are very similar to these.)
The uncorrected proton structure functions $F_i^{(0)}$ are
constructed from the CTEQ-Jefferson Lab (CJ) global PDF fits
\cite{CJ11}, evaluated at $Q^2 = 2$~GeV$^2$.  For each of the
structure functions the effects of TMCs become more prominent
with increasing~$x$, and naturally their magnitude decreases
at larger $Q^2$.

For the $F_1^\gamma$ structure function in Fig.~\ref{fig:F1},
the deviation from unity of the ratio of target mass corrected
to uncorrected functions ranges from $\sim 10\%$ at $x = 0.4$,
to more than a factor 2 for $x \gtrsim 0.7$.
The model dependence of the TMCs to $F_1^\gamma$ is relatively weak;
the OPE \cite{GP76} and EFP \cite{EFP83} results are similar to within
a few percent for all $x$, while the $\xi$-scaling \cite{AOT94,KR02}
and AQ \cite{AQ08} prescriptions differ from the OPE by $\lesssim 5\%$
and 15\%, respectively, for $x \lesssim 0.8$.
(Results for $x \gtrsim 0.9$ are not shown as the input nucleon PDFs
are not constrained in this region and display numerical instability
at $x \gtrsim 0.95$.)
In fact, at low and moderate $x$ the OPE and EFP TMCs track each other
rather closely, as expected from the equality of their leading term
prefactors at order $1/Q^2$.  Similarly, the AQ and $\xi$-scaling
prescriptions are much closer to each other than to the OPE and EFP
results, as may be anticipated from the structure of the respective
TMC expressions in Eqs.~(\ref{eq:AQ}) and (\ref{eq:xiS}).

Qualitatively similar behavior is seen also for the target mass
corrected $F_2^\gamma$ structure function in Fig.~\ref{fig:F2}.
Here a dip in the ratio of corrected to uncorrected functions
at $x \sim 0.4$, however, delays the sharp rise above unity to
$x \gtrsim 0.6$.  As for $F_1^\gamma$, the EFP result agrees with
the OPE to a few percent over the entire $x$ range, and the AQ
and $\xi$-scaling ratios are almost identical for $x < 0.4$.
The two sets of ratios differ by $\lesssim 7\%$ for $x < 0.7$,
before diverging somewhat as $x \to 1$.

For the $F_L^\gamma$ structure function in Fig.~\ref{fig:FL} the
differences between the various TMC prescriptions are more dramatic.
The OPE and EFP ratios begin to rise steeply at low $x$, with the
corrected functions exceeding the uncorrected ones by more than a
factor 5 already by $x = 0.5$.  The AQ and $\xi$-S ratios, on the
other hand, rise above unity at much higher $x$, reminiscent of the
$F_1$ ratios in Fig.~\ref{fig:F1}.  The two sets of corrections differ
by more than 50\% by $x=0.3$, and by $\gtrsim 80\%$ for $x > 0.8$.
This qualitatively rather different behavior can be understood
by directly comparing Eqs.~(\ref{eq:OPE}) and (\ref{eq:EFP}) to
Eqs.~(\ref{eq:AQ}) and (\ref{eq:xiS}).  Unlike the $\xi$-S and AQ
prescriptions, the OPE and EFP $F_L$ results include terms involving
integrals over $F_2$, which is generally $\gg F_L$.  In fact, to
leading order with no TMCs, the $F_L$ structure function vanishes,
and adding NLO corrections within the AQ or $\xi$-S prescriptions
does not produce a significant increase.  In contrast, the OPE and
EFP prescriptions always receive large $F_2$ contributions, making
the target mass corrected to uncorrected ratio considerably larger
in these approaches.

The strong correlations between the OPE and EFP predictions are
not as visible for the $F_3$ structure function, which, unlike
the other structure functions, differs already at ${\cal O}(1/Q^2)$.
The general shape of the TMC ratio, illustrated in Fig.~\ref{fig:F3}
for the $F_3^{W^+}$ structure function, resembles that for $F_2^\gamma$
in Fig.~\ref{fig:F2}, but with a rise above unity beginning at lower $x$.
The various prescriptions agree to $\sim 10\%$ for $x \lesssim 0.4$,
and $\sim 40\%$ for $x \lesssim 0.8$, but generally display more spread
than in $F_1^\gamma$ or $F_2^\gamma$.

\newpage
\section{Implications for Observables}
\label{sec:obs}

Having examined the differences between the various TMC prescriptions
on individual structure functions, in this section we consider the
effects of TMCs, and in particular their model dependence, on several
observables that will be measured in upcoming experiments.
These include the ratio of the neutron to proton $F_2$ structure
functions, the longitudinal to transverse (LT) cross section ratios,
and parity-violating deep-inelastic scattering asymmetries for the
proton and deuteron.

\subsection{Neutron to proton ratio}
\label{ssec:np}

Historically, the ratio of $d$ to $u$ quark distributions in the proton
has been determined primarily through the ratio of neutron to proton
electromagnetic DIS cross sections,
\begin{eqnarray}
R_{np}\
 =\ { d^2\sigma^{\gamma n}/dxdy \over d^2\sigma^{\gamma p}/dxdy }
&=& { F_2^{\gamma n} \over F_2^{\gamma p} }
\left(
{ 1 - y - y^2 \left[ \rho^2-1 - 2\rho^2/(1+R^{\gamma n}) \right]/4
  \over
  1 - y - y^2 \left[ \rho^2-1 - 2\rho^2/(1+R^{\gamma p}) \right]/4 }
\right),
\label{eq:Rnp}
\end{eqnarray}
where $y = \nu/E$ is the fractional energy transfer from the lepton,
and $R^{\gamma N}$ is the ratio of the longitudinal to transverse
cross sections, or structure functions,
\begin{eqnarray}
\label{eq:R_LT}
R^{\gamma N} &=& { F_L^{\gamma N} \over 2x F_1^{\gamma N} }
\end{eqnarray}
for nucleon $N$.
With the assumption that $R^{\gamma n} = R^{\gamma p}$, the ratio
of cross sections becomes the ratio of $F_2$ structure functions,
$R_{np} \to F_2^{\gamma n}/F_2^{\gamma p}$.
To leading order, the ratio (\ref{eq:Rnp}) is then given by
$R_{np} = (1 + 4 d/u)/(4 + d/u)$, which illustrates the sensitivity
to the $d/u$ PDF ratio.
In practice, differences between $R^{\gamma p}$ and $R^{\gamma n}$
generated perturbatively at NLO have a negligible effect on the
ratio $R_{np}$ at the kinematics considered here.

The absence of free neutron targets has meant that in practice
inclusive deuterium structure function data has been used to obtain
indirect information on the neutron, and hence the $d$ quark.
This procedure is known to suffer from significant model dependence
at large values of $x$ \cite{CJ11, MT96, Afnan00}, leading to several
novel new experiments being proposed \cite{BONUS12, MARATHON, SOLID}
to determine the $d/u$ ratio with minimal nuclear model uncertainties.
In order for these measurements to be unambiguously analyzed,
it is important to quantify the extent of TMC uncertainties at
the kinematics of the experiments, which will typically reach a
maximum $Q^2 \sim 10 - 15$~GeV$^2$ at $x \approx 0.8$.

\begin{figure}[ht]
\rotatebox{-90}{\includegraphics[width=5.7cm]{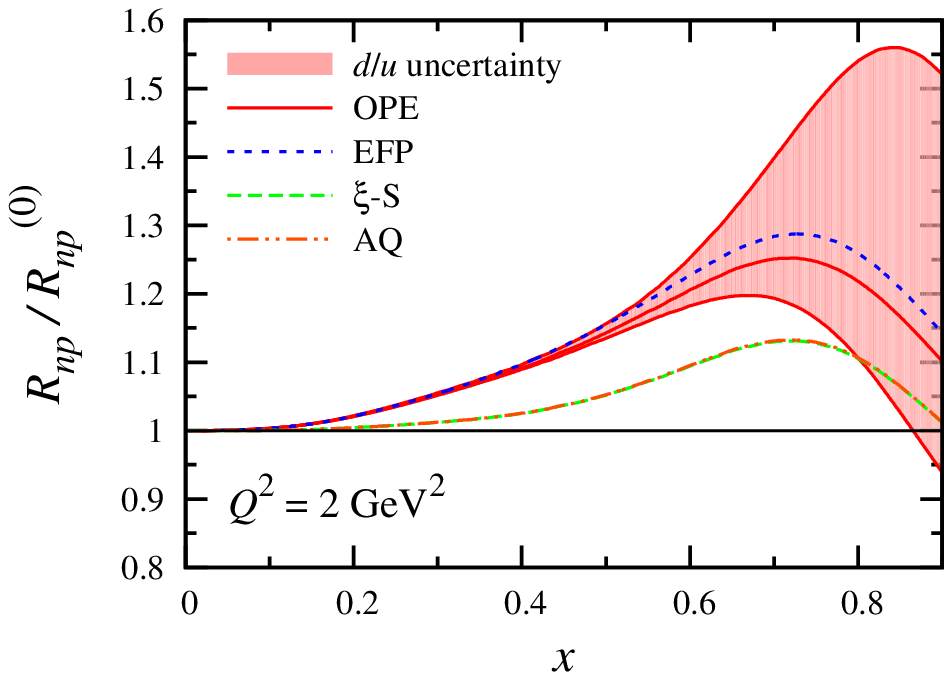}}
\rotatebox{-90}{\includegraphics[width=5.7cm]{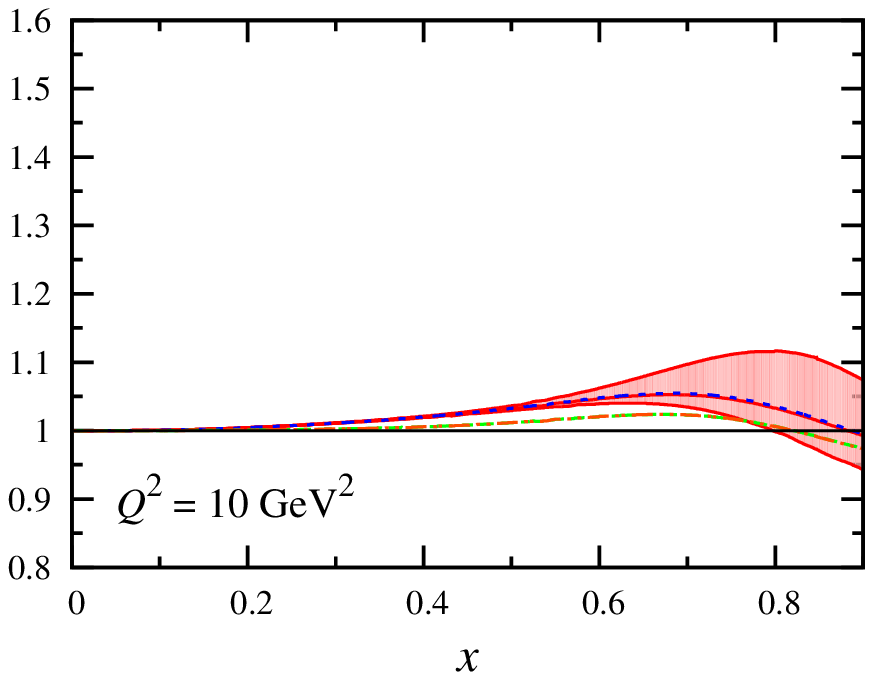}}
\caption{Ratio of target mass corrected ($R_{np}$) to uncorrected
	($R_{np}^{(0)}$) neutron to proton $F_2$ structure function
	ratios at $Q^2 = 2$~GeV$^2$ {\bf (left)} and $Q^2 = 10$~GeV$^2$
	{\bf (right)},
	for the OPE (solid, red),
	EFP (short-dashed, blue),
	$\xi\mbox{-S}$ (long-dashed, green), and
	AQ (dot-dashed, orange) TMC prescriptions.
	The shaded band represents the uncertainty in the ratio $d/u$ for
	the OPE result.  Note that the AQ and $\xi\mbox{-S}$ results are
	almost identical.}
\label{fig:NP}
\end{figure}

The effects of TMCs on the ratio $R_{np}$ are illustrated in
Fig.~\ref{fig:NP} for the CJ PDFs \cite{CJ11} at $Q^2 = 2$~GeV$^2$
(left) and 10~GeV$^2$ (right).  The shaded bands represent the $d/u$
uncertainty range as applied to the OPE TMC prescription, with the
central solid (red) curve denoting the median value for OPE
calculated with the same PDFs as the other TMC prescriptions.
The target mass corrections at $Q^2=2$~GeV$^2$ are sizable, reaching
$\approx 25\% - 30\%$ at $x = 0.7$ for the OPE and EFP prescriptions,
and $\approx 12\%$ for the $\xi$-S and AQ results.  At the higher
$Q^2=10$~GeV$^2$ value the TMCs decrease to $\approx 5\%$ and
$\approx 2\%$ for the OPE/EFP and $\xi$-S/AQ calculations,
respectively.  Treating each of the TMC prescriptions on equal
footing, this would suggest an uncertainty due to TMCs of
$\lesssim 3\%$ for all values of $x$ accessible in the planned
experiments \cite{BONUS12, MARATHON, SOLID}.

The TMC uncertainty can be compared with the range of $R_{np}$
predicted from PDFs extracted under different assumptions about
the size of nuclear corrections in deuterium, which currently
represents the largest uncertainty in the $d/u$ ratio at
$x \gtrsim 0.5$ \cite{CJ11}.  This is illustrated in the bands
in Fig.~\ref{fig:NP}, which represent the $R_{np}$ ratio evaluated 
from the range of CJ PDFs \cite{CJ11} for the OPE TMC prescription.
The results show that for $Q^2 = 2$~GeV$^2$ the uncertainty resulting
from nuclear corrections is some $2-3$ times larger than that
associated with TMCs at $x=0.8$.  Both the uncertainties in the $d$
quark PDF and in the TMCs decrease as $x$ decreases, albeit more
slowly for the latter.  At $x=0.6$, in fact, the two uncertainties
are comparable, while for $x \lesssim 0.4$ the TMC uncertainty is
actually larger.

\begin{figure}[t]
\rotatebox{-90}{\includegraphics[width=5.7cm]{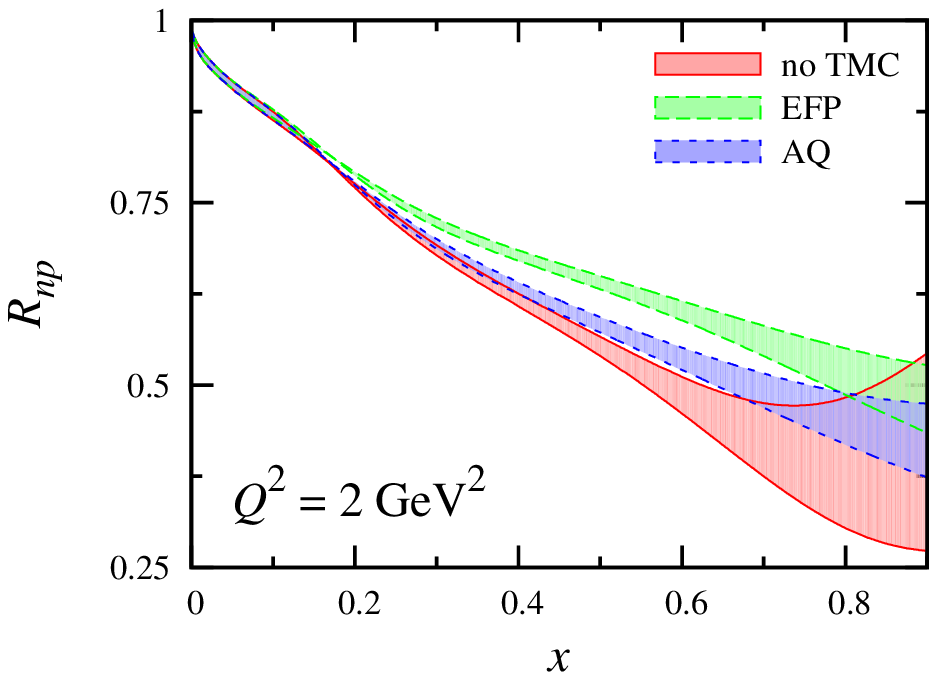}}
\rotatebox{-90}{\includegraphics[width=5.7cm]{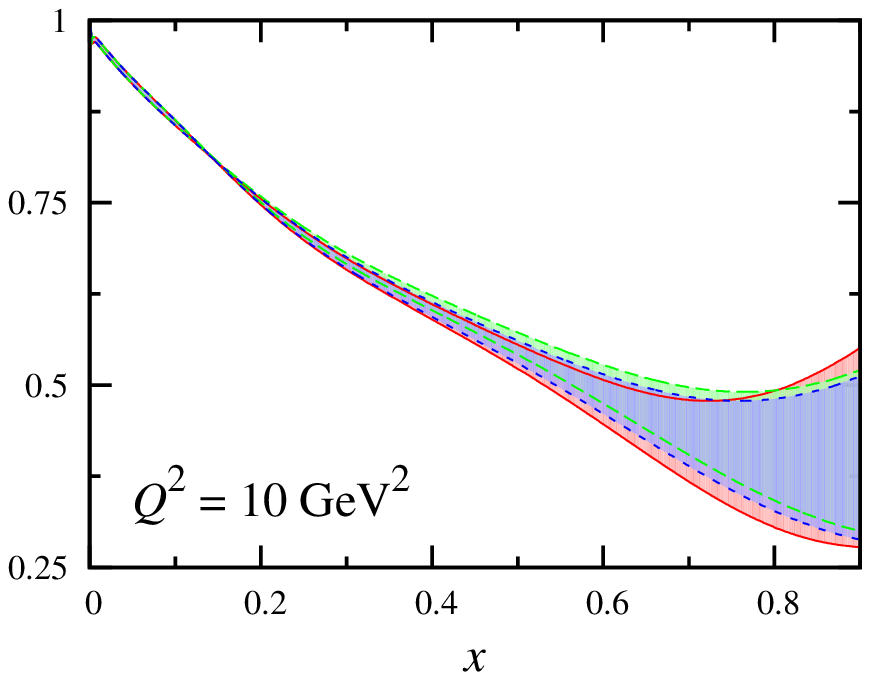}}
\caption{Neutron to proton $F_2$ structure function ratio $R_{np}$
	at $Q^2 = 2$~GeV$^2$ {\bf (left)} and $Q^2 = 10$~GeV$^2$
	{\bf (right)}, with the shaded bands representing the $d/u$
	uncertainty for uncorrected structure functions (solid, red)
	and the two extremal TMCs, EFP (long-dashed, green)
	and AQ (short-dashed, blue).}
\label{fig:NP_test}
\end{figure}

With increasing $Q^2$ both the TMCs and their uncertainties decrease,
while the uncertainty in the leading twist PDFs remains approximately
$Q^2$ independent, as can be seen more clearly in the comparison of
the absolute values of $R_{np}$ in Fig.~\ref{fig:NP_test}.
At $Q^2=2$~GeV$^2$ the bands representing the $R_{np}$ ratio evaluated
from the CJ PDFs \cite{CJ11} using different TMC prescriptions
(specifically the extremal EFP and AQ results) do not overlap until
$x \sim 0.75$, meaning that at smaller $x$ the true $d/u$ behavior
will be obscured by the relatively large TMC model uncertainty.
Interestingly, the TMCs actually {\it decrease} the nuclear uncertainty
range at lower $Q^2$, since the action of the $x \to \xi$ rescaling
is to effectively feed information from lower $x$ in the uncorrected
functions (which have relatively small $d/u$ uncertainty) to higher
$x$ values (where the $d/u$ uncertainty is larger).
Consequently, at higher $Q^2$ the sensitivity to the $d/u$ ratio
increases both due to the smaller spread of results for different
TMC prescriptions, and to the weakening of the TMC effect in moving
strength from lower $x$ for a particular TMC prescription.
This is indeed visible in Fig.~\ref{fig:NP_test} for the $R_{np}$
ratio at $Q^2=10$~GeV$^2$, in which the EFP and AQ extremal TMC
bands very nearly coincide over the entire $x$ range, as well as
with the ratio computed without TMCs.  Such values of $Q^2$ will
therefore be required in order to cleanly extricate the $d/u$ ratio
from measurements of the neutron to proton ratio without ambiguities
associated with TMC model dependence.

Finally, we also note that in global QCD fits of PDFs it was
recently found \cite{CTEQ6X} that the impact of the model dependence
of TMCs on leading twist PDFs is reduced significantly with the
inclusion of a phenomenological $1/Q^2$ higher twist term in the
$F_2$ structure function parametrization, with the two effects
partially compensating each other.  Since the higher twist contribution
to $F_L$ is independent of that for $F_2$, a similar cancellation may
be expected also when including $F_L$ data in the global fits.
The uncertainties in extracted PDFs induced by the model dependence of
TMCs may therefore be smaller than those suggested in Fig.~\ref{fig:NP}
if the data are analyzed within a global PDF context.

\subsection{Longitudinal to transverse structure function ratios}
\label{sec:LTRat}

While the LT cross section ratio $R^{\gamma N}$ is expected to play
a relatively minor role in the measurements of the neutron to proton
$F_2$ structure function ratio in Eq.~(\ref{eq:Rnp}), mostly because of
the cancellation between the proton and neutron $R^{\gamma N}$ values,
the effects of TMCs on the ratio itself may be more significant.
This was already suggested by the large prescription dependence of TMCs
for the longitudinal structure function $F_L^\gamma$ in Fig.~\ref{fig:FL}.
The effects of TMCs on the LT ratio are also important to quantify in
connection with establishing the low-$Q^2$ behavior of $R^{\gamma N}$
at finite $x$, to determine the onset of gauge invariance constraints
on the longitudinal structure function \cite{MEK05}.

\begin{figure}[htb]
\rotatebox{-90}{\includegraphics[width=5.7cm]{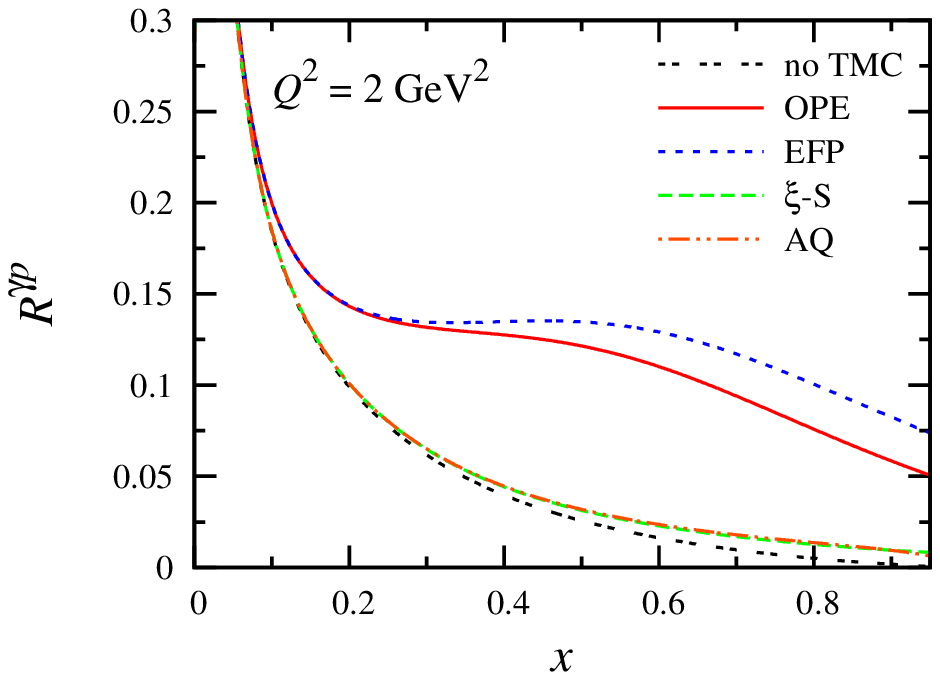}}
\rotatebox{-90}{\includegraphics[width=5.7cm]{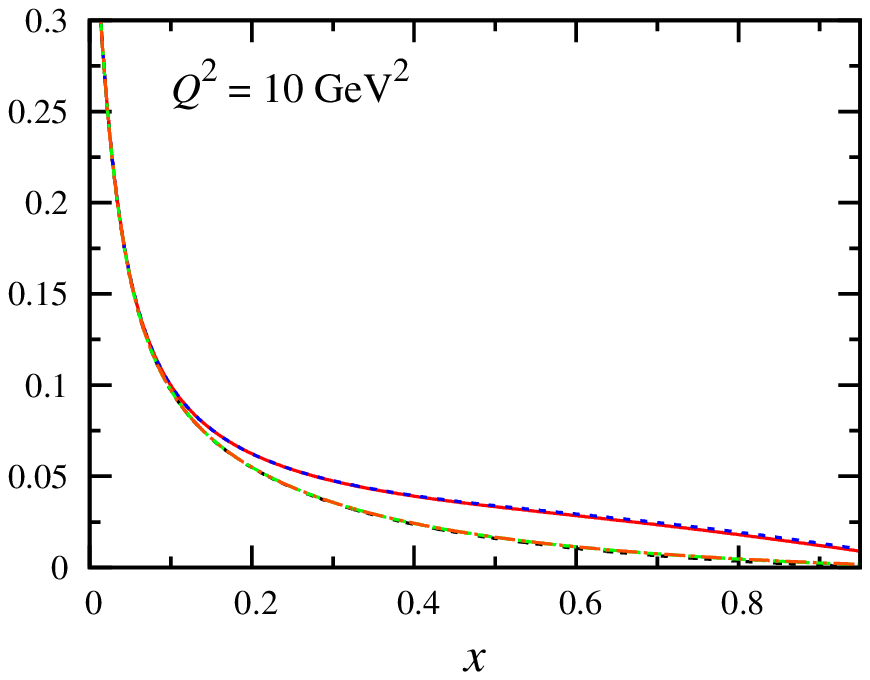}}
\caption{Longitudinal to transverse cross section ratio $R^{\gamma p}$
	for the proton at $Q^2 = 2$~GeV$^2$ {\bf (left)} and 
	$Q^2 = 10$~GeV$^2$ {\bf (right)}, for
	no TMCs (double-dashed, black),
	the OPE (solid, red),
	EFP (short-dashed, blue),
	$\xi$-S (long-dashed, green), and
	AQ (dot-dashed, orange) TMC prescriptions.}
\label{fig:R_LT}
\end{figure}

In Fig.~\ref{fig:R_LT} we illustrate the TMC effects on $R^{\gamma p}$
for $Q^2 = 2$ and 10~GeV$^2$ for each of the TMC prescriptions considered.
All of the TMCs increase the magnitude of the $R^{\gamma p}$ ratio, with
the AQ and $\xi$-S prescriptions having a relatively modest effect
(approximately a factor 2 for $x \approx 0.6-0.8$ at $Q^2 = 2$~GeV$^2$,
but only a few percent at $Q^2 = 10$~GeV$^2$), while the EFP and OPE both
altering the ratio significantly for $x \gtrsim 0.1$.  The enhancement
of $R^{\gamma p}$ for the latter is predicted to be about an order of
magnitude for $x \approx 0.6-0.8$ at $Q^2 = 2$~GeV$^2$, and still a
factor of $3-4$ at $Q^2 = 10$~GeV$^2$.

Some differences are also expected between the longitudinal to
transverse cross section ratios at NLO for processes involving
electromagnetic and weak currents.  In particular, as will be
discussed in more detail in Sec.~\ref{sec:PVDIS} below, asymmetries
measured in parity-violating electron scattering are sensitive
to interference effects between $\gamma$ and $Z$ boson exchange,
and differences between the $R^\gamma$ and $R^{\gamma Z}$ LT ratios
can affect the measured asymmetries~\cite{Hobbs08, Hobbs11}.

\begin{figure}[htb]
\rotatebox{-90}{\includegraphics[width=5.7cm]{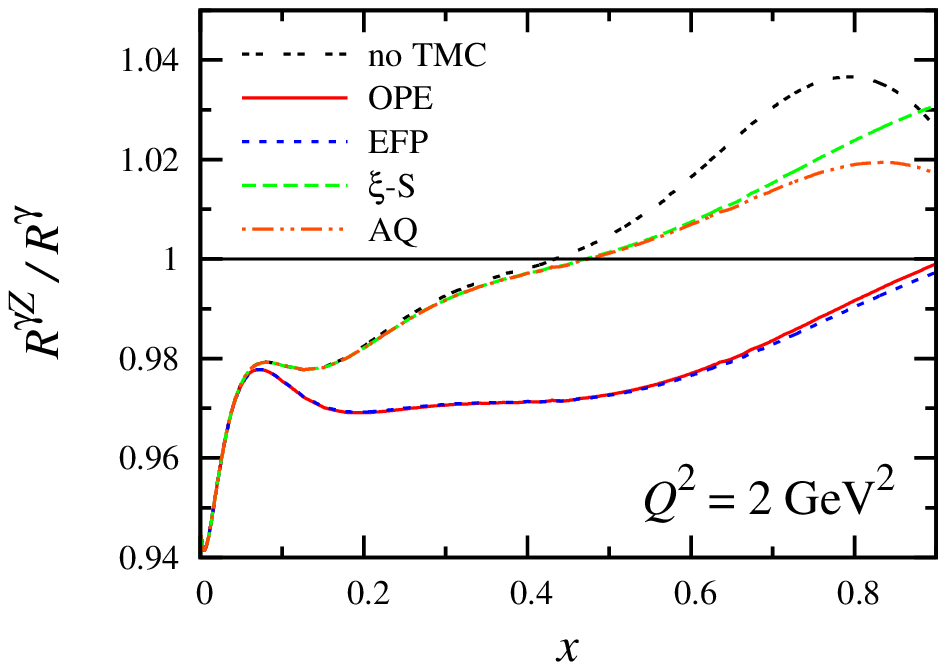}}
\rotatebox{-90}{\includegraphics[width=5.7cm]{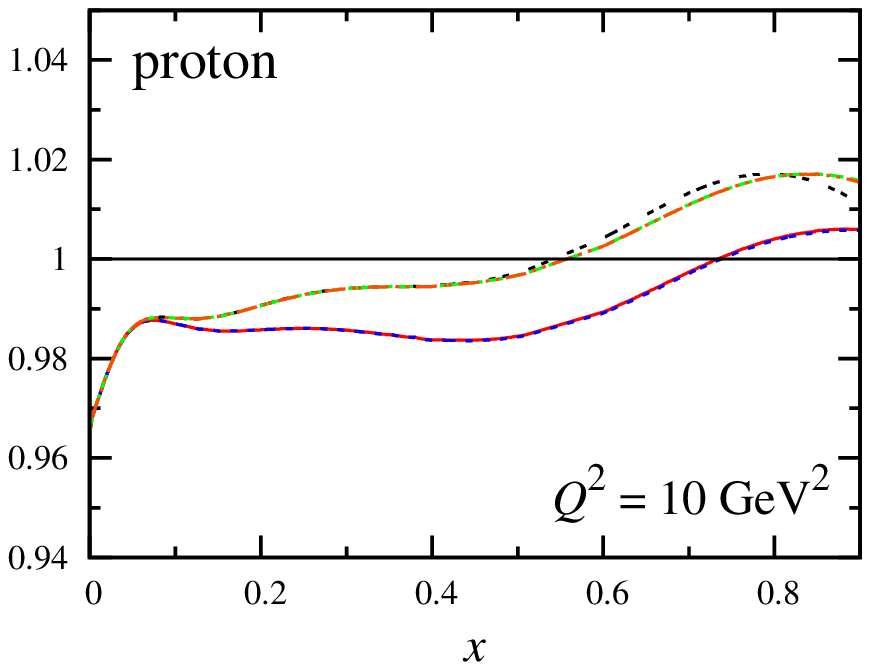}}
\caption{Ratio of $R^{\gamma Z}$ to $R^{\gamma}$ LT ratios for the
	{\it proton} computed at NLO for $Q^2 = 2$~GeV$^2$ {\bf (left)}
	and $Q^2 = 10$~GeV$^2$ {\bf (right)}, for
	no TMCs (double-dashed, black),
	the OPE (solid, red),
	EFP (short-dashed, blue),
	$\xi$-S (long-dashed, green), and
	AQ (dot-dashed, orange) TMC prescriptions.}
\label{fig:Rp}
\end{figure}

In Fig.~\ref{fig:Rp} the ratio of the proton $R^{\gamma Z}$
to $R^\gamma$ LT ratios is shown at $Q^2 = 2$ and 10~GeV$^2$.
While at leading order both of these ratios are zero, at NLO
the different relative contributions from quark PDFs to the
electromagnetic and $\gamma Z$ interference structure functions
leads to deviations of the ratios from unity of up to $\approx 4\%$
at $Q^2 = 2$~GeV$^2$, and up to $\approx 2\%$ at $Q^2 = 10$~GeV$^2$.
The effects of the TMCs are again very small for the $\xi$-scaling
and AQ prescriptions, but more significant for the OPE and EFP
results.  Overall, the spread in the TMC predictions for the
$R^{\gamma Z}/R^\gamma$ ratio amounts to $\lesssim 4 - 5\%$ for
$x$ between 0.6 and 0.8 at $Q^2 = 2$~GeV$^2$, and $\lesssim 2\%$
at $Q^2 = 10$~GeV$^2$.
Note that the dip in the ratios at $x < 0.1$, which is insensitive
to TMCs, reflects the greater role played by gluons at low $x$,
but is mostly irrelevant for the kinematics of the proposed
experiments \cite{BONUS12, MARATHON, SOLID}.

\begin{figure}[htb]
\rotatebox{-90}{\includegraphics[width=5.7cm]{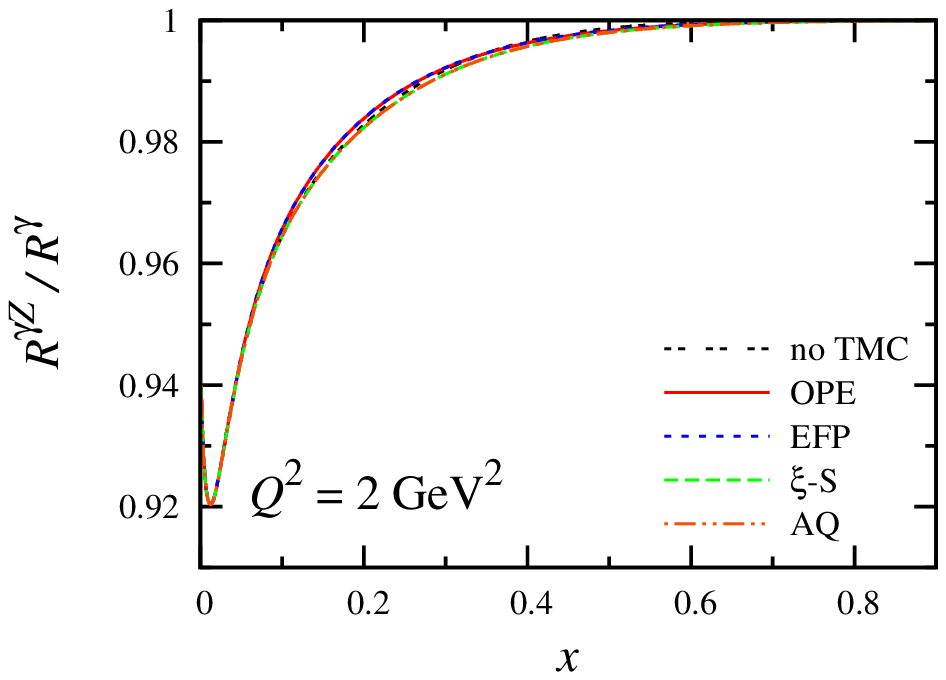}}
\rotatebox{-90}{\includegraphics[width=5.7cm]{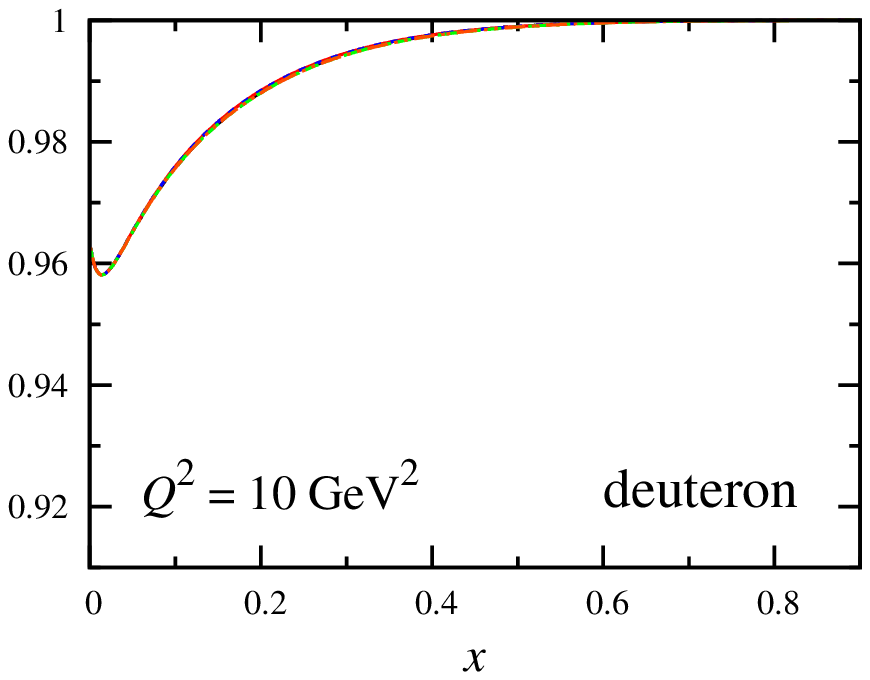}}
\caption{As in Fig.~\ref{fig:Rp}, but for the {\it deuteron}
	$R^{\gamma Z}$ to $R^{\gamma}$ LT ratio.}
\label{fig:Rd}
\end{figure}

For the case of the isoscalar deuteron target, stronger cancellations
between the quark content of $R^{\gamma Z}$ and $R^{\gamma}$ are expected
to lead to smaller deviations of their ratio from unity at large $x$.
This is indeed observed in Fig.~\ref{fig:Rd}, where again the dip in the
ratio at very low $x$ is associated with NLO gluon dominance of the LT
ratios as $x \to 0$.  At $x=0.2$, for example, the gluonic content of
$F_L$ suppresses the deuteron $R^{\gamma Z}/R^{\gamma}$ ratio by
$\approx 2\%$ for $Q^2 = 2$~GeV$^2$, and $\approx 1\%$ for
$Q^2 = 10$~GeV$^2$.
At higher $x$ the deviations decrease until the ratio approaches unity
asymptotically as $x \to 1$.  In the region of $x$ where the LT ratios
are dominated by quarks, the fact that the same isoscalar combination of
quark PDFs enters both the electromagnetic and $\gamma Z$ interference
structure functions leads to almost negligible TMC effects.
The absence of significant TMC effects in the deuteron ratio is,
as expected, even more clearly visible at the higher $Q^2$ value.

\begin{figure}[t]
\rotatebox{-90}{\includegraphics[width=7cm]{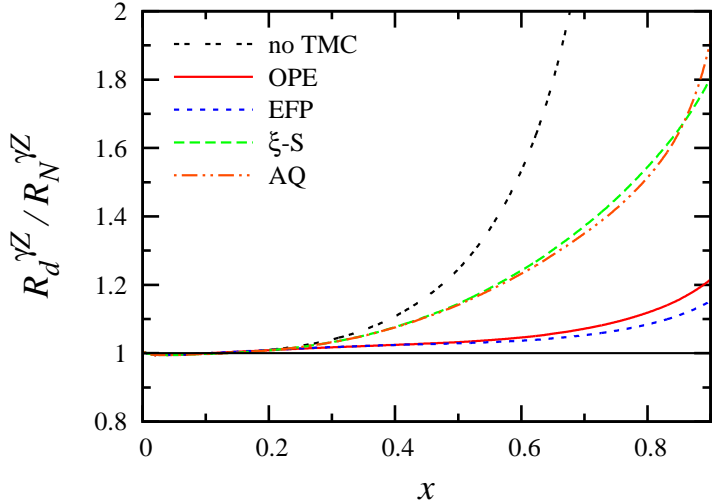}}
\caption{Ratio of the NLO $R^{\gamma Z}$ LT ratios for the deuteron
	and isoscalar nucleon ($N = p+n$) at $Q^2 = 2$~GeV$^2$ for
	no TMCs (double-dashed, black) and the 
	OPE (solid, red),
	EFP (short-dashed, blue),
	$\xi$-S (long-dashed, green), and
	AQ (dot-dashed, orange) TMC prescriptions.}
\label{fig:smear1}
\end{figure}

Finally, for experiments involving deuteron targets one needs to
account for the fact that the nucleons in the nucleus are bound and
hence have structure functions that differ from those of free nucleons.
In Fig.~\ref{fig:smear1} the LT ratio $R^{\gamma Z}$ for the $\gamma Z$
interference structure functions of the deuteron is shown relative to
that for a free isoscalar nucleon (defined as proton $+$ neutron) at
$Q^2 = 2$~GeV$^2$, using the nuclear smearing model of
Refs.~\cite{KP06,KMK09} (see also Ref.~\cite{CJ11}).  
The shape of the deuteron to nucleon ratio computed at NLO in the
absence of TMCs displays a dramatic rise above unity with increasing
$x$ that is characteristic of the nuclear EMC ratio \cite{EMC}.
The effects of Fermi motion in fact lead to a divergent ratio at $x=1$.
The inclusion of TMCs suppresses the rise at large $x$, from
$\approx 60\%$ at $x=0.6$ to 20\% for the $\xi$-scaling and AQ
prescriptions, and to $\approx 5\%$ for the OPE and EFP cases,
with larger differences as $x \to 1$.  This suppression arises because
TMCs shift strength in $F_L$ from small $x$ to large $x$, thereby
lessening the impact of the nuclear smearing.  Since the TMC effects
in the OPE and EFP prescriptions are larger than that for the $\xi$-S
and AQ cases (see Fig.~\ref{fig:FL}), the nuclear corrections for the
former in Fig.~\ref{fig:smear1} are smaller.

The smearing corrections to the electromagnetic LT ratio $R^{\gamma}$
are almost identical to those in Fig.~\ref{fig:smear1}.  Consequently,
the net effect on the ratio $R^{\gamma Z}/R^{\gamma}$ for the deuteron
computed with or without nuclear corrections is $\lesssim 0.05\%$, and
can be neglected for the kinematics of interest.

\subsection{Parity-violating DIS}
\label{sec:PVDIS}

Measurements of parity-violating deep-inelastic scattering (PVDIS)
asymmetries on the proton have been proposed at Jefferson Lab
\cite{SOLID} to provide independent constraints on the ratio of $d/u$
quark distributions at large $x$, free of the nuclear correction
uncertainties associated with deuterium measurements \cite{PVDIS-p}.
In the case of deuteron targets, where much of the dependence on
hadron structure effects cancels, PVDIS asymmetries are sensitive
to several effects, including charge symmetry violation in PDFs
\cite{CSVrev}, or to Standard Model parameters whose precise
measurement can reveal signals for new physics \cite{Bj78,PVDIS-SLAC}.
In this section we examine the effects of TMCs on the PVDIS asymmetries
of the proton and deuteron, and discuss the phenomenological
implications of their uncertainties on future planned experiments.

The PV asymmetry is defined through the difference and sum of the
inclusive cross sections for scattering either a right-handed ($R$)
or left-handed ($L$) electron from an unpolarized target,
\begin{equation}
A_{\rm PV} = \frac{\sigma_R - \sigma_L}{\sigma_R + \sigma_L},
\end{equation}
where $\sigma_{R,L} \equiv (d^2\sigma/d\Omega dE')_{R,L}$.
Since the purely electromagnetic and purely weak contributions to the
cross section are independent of electron helicity for $Q^2 \ll M_Z^2$,
they cancel in the numerator, leaving only the $\gamma Z$ interference
term.  The denominator, on the other hand, contains all contributions,
but is dominated by the purely electromagnetic component.
In terms of structure functions, the asymmetry can be written
\cite{Hobbs08}
\begin{equation}
A_{\rm PV}
= - \left( {G_F Q^2 \over 2 \sqrt{2} \pi \alpha} \right)
  \left[ g_A^e\ Y_1\ \frac{F_1^{\gamma Z}}{F_1^\gamma}\
     +\ {g_V^e \over 2}\ Y_3\ {F_3^{\gamma Z} \over F_1^\gamma} 
  \right],
\label{eq:APV}
\end{equation}
where $g_A^e = -1/2$ and $g_V^e = -1/2 + 2 \sin^2\theta_W$ are the
axial and vector couplings of the $Z$ boson to the electron, with
$\theta_W$ the weak mixing angle, and the functions $Y_{1,3}$ parametrize 
the dependence on $y$ and on the $R^\gamma$ and $R^{\gamma Z}$ ratios,
\begin{subequations}
\label{eq:Y}
\begin{align}
\label{eq:Y1}
Y_1
&={ 1 + (1-y)^2 - y^2 [1 + \rho^2 - 2\rho^2/(R^{\gamma Z}+1)] / 2
  \over
    1 + (1-y)^2 - y^2 [1 + \rho^2 - 2\rho^2/(R^\gamma+1)] / 2 }
  \left( {1+R^{\gamma Z} \over 1+R^\gamma} \right),		\\
\label{eq:Y3}
Y_3
&={ 1 - (1-y)^2 
  \over
    1 + (1-y)^2 - y^2 [1 + \rho^2 - 2\rho^2/(R^\gamma+1)] / 2 }
   \left( {\rho^2 \over 1+R^\gamma} \right).
\end{align}
\end{subequations}
In the limit of $Q^2 \to \infty$, where $\rho \to 1$ and
$R^{\gamma,\gamma Z} \to 0$, the kinematical factors simplify to
$Y_1 \to 1$ and $Y_3 \to [1 - (1-y)^2]/[1 + (1-y)^2]$.

\subsubsection{Proton asymmetry}
\label{ssec:PVDIS-p}

\begin{figure}[h]
\rotatebox{-90}{\includegraphics[width=5.7cm]{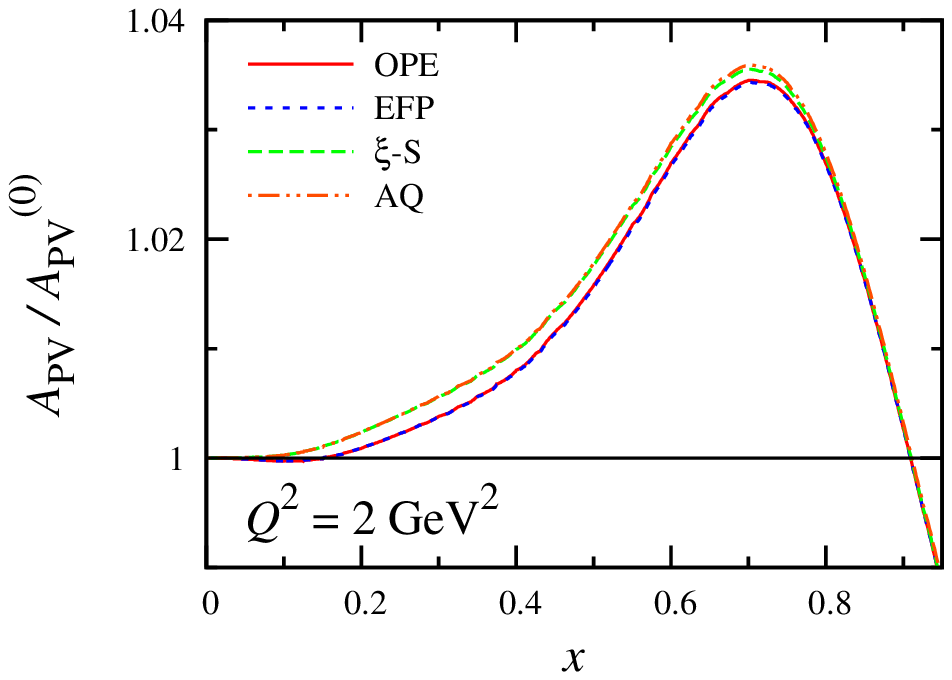}}
\rotatebox{-90}{\includegraphics[width=5.7cm]{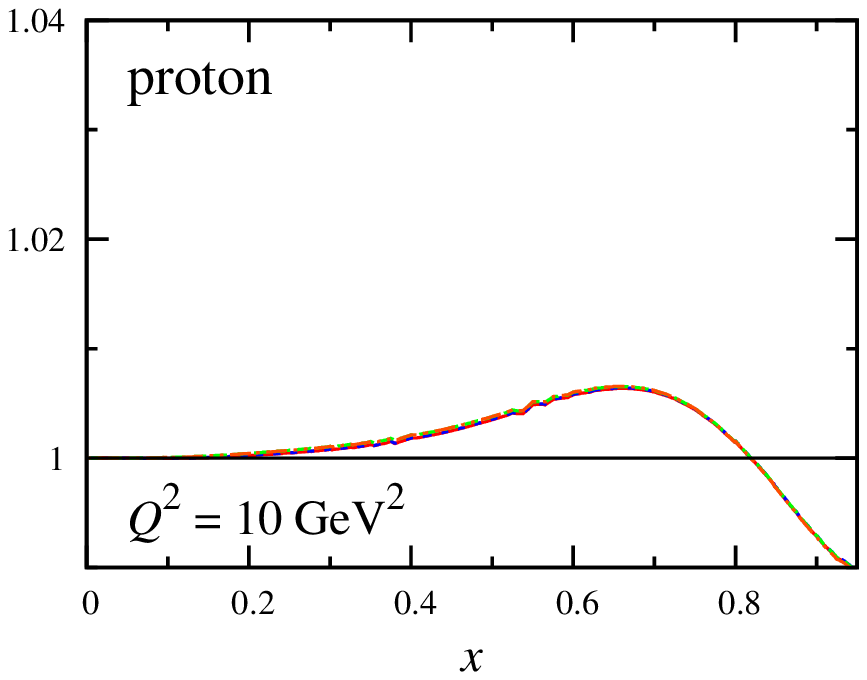}}
\caption{Ratio of target mass corrected ($A_{\rm PV}$) to uncorrected
	($A_{\rm PV}^{(0)}$) PVDIS asymmetries for the {\it proton}
	at $Q^2 = 2$~GeV$^2$ {\bf (left)} and $Q^2 = 10$~GeV$^2$
	{\bf (right)}, for the	OPE (solid, red),
	EFP (short-dashed, blue),
	$\xi$-S (long-dashed, green), and
	AQ (dot-dashed, orange) TMC prescriptions.
	Note that the AQ and $\xi$-S results are almost
	indistinguishable, as are the EFP and OPE prescriptions.}
\label{fig:PAS}
\end{figure}

The proton PVDIS asymmetry is shown in Fig.~\ref{fig:PAS} for
$Q^2=2$ and 10~GeV$^2$ in the form of the ratio of the target mass
corrected to uncorrected asymmetries.  For all prescriptions the TMC
effects are maximal at $x \approx 0.7$, where they are of the order
of $3-4\%$ at $Q^2=2$~GeV$^2$ and $\lesssim 1\%$ at $Q^2=10$~GeV$^2$.
The results are slightly higher for the $\xi$-S and AQ corrections
(which are virtually indistinguishable) than for the OPE and EFP
(which are also almost identical).  The small size of the effects is
principally due to the strong cancellation of the TMCs in the $F_1$
structure functions, namely,
$( F_1^{\gamma Z} / F_1^\gamma )^{\rm TMC}
 \approx 
 ( F_1^{\gamma Z} / F_1^\gamma )^{(0)}$,
even though $|F_1^{\rm TMC} / F_1^{(0)}| \gg 1$ at high $x$. 
Overall, the results indicate that the asymmetries themselves are
less sensitive to TMCs than are the LT ratios $R^{\gamma, \gamma Z}$
on which the asymmetries depend.

\begin{figure}[t]
\rotatebox{-90}{\includegraphics[width=5.7cm]{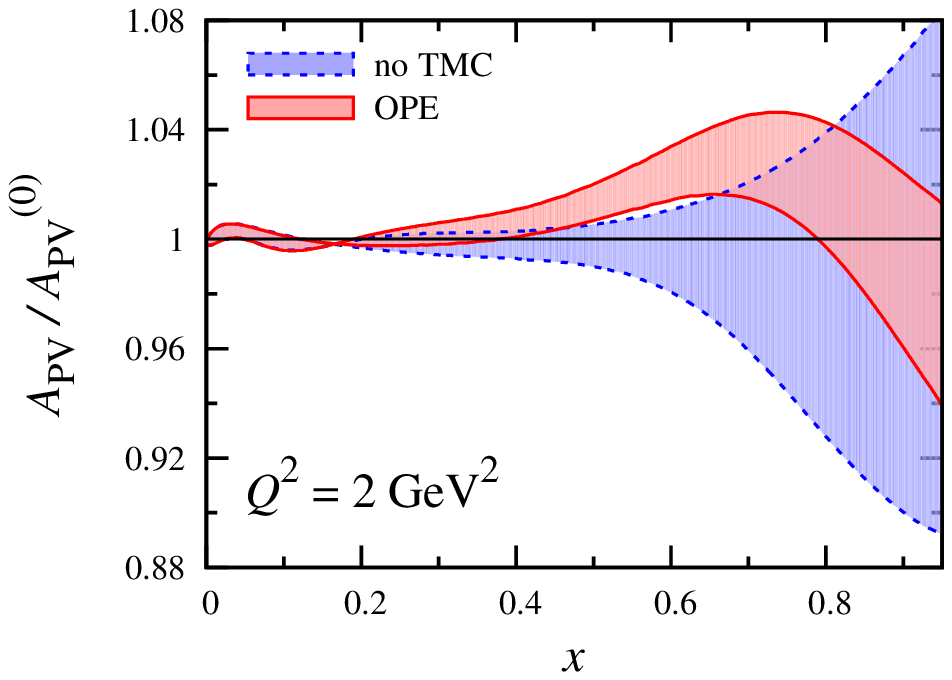}}
\rotatebox{-90}{\includegraphics[width=5.7cm]{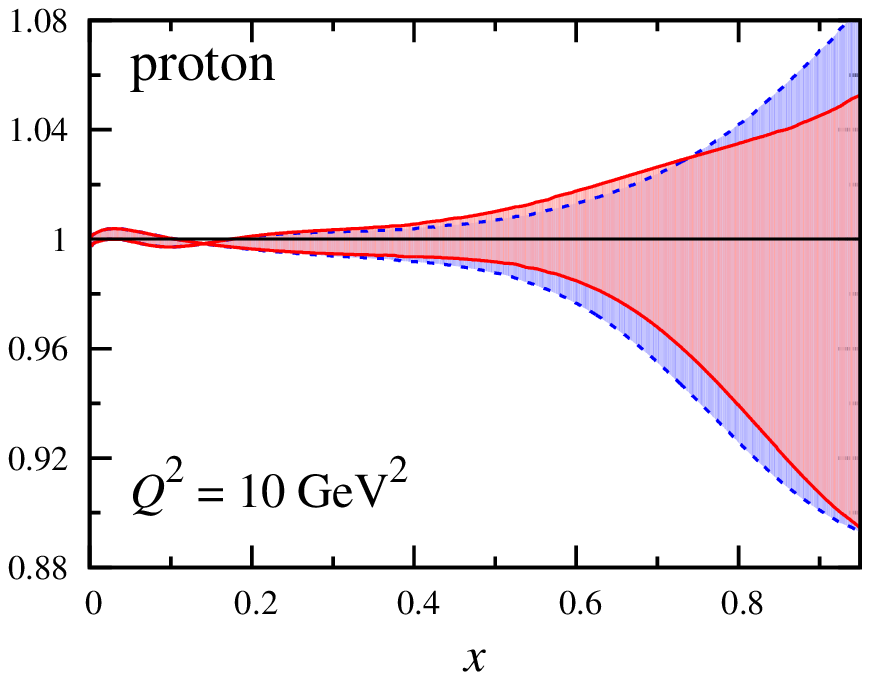}}
\caption{Proton PVDIS asymmetry $A_{\rm PV}$ at $Q^2 = 2$~GeV$^2$
	{\bf (left)} and $Q^2 = 10$~GeV$^2$ {\bf (right)} for
	CJ PDFs with minimal and maximal nuclear effects \cite{CJ11}
	(shaded bands), relative to the asymmetry $A_{\rm PV}^{(0)}$
	for the central CJ PDF fits, with no TMCs (dashed, blue) and
	using the OPE TMC prescription (solid, red).}
\label{fig:PASOPE}
\end{figure}

Since one of the main goals of the proton PVDIS measurements will be
to reduce the uncertainties on PDFs at large $x$, particularly on the
$d/u$ ratio, it is instructive to compare the magnitude of the TMC
effects with the expected sensitivity of the asymmetry to different
possible PDF behaviors at large $x$.  In Fig.~\ref{fig:PASOPE} we show
the proton asymmetry $A_{\rm PV}$ computed from the full range of CJ
PDFs \cite{CJ11} including minimal and maximal nuclear corrections
(shaded bands) relative to the central PDF fits.  The uncertainty band
increases with increasing $x$, reflecting the larger uncertainty on
the $d$ quark PDF at large $x$, and in the absence of TMCs ranges from
$\approx 3\%$ at $x=0.6$ to $\approx 11\%$ at $x=0.8$ for both $Q^2=2$
and 10~GeV$^2$.  This is significantly larger than the TMC uncertainty
band in Fig.~\ref{fig:PAS}, where the spread of the TMC model
predictions is $\ll 1\%$, even though the absolute target mass
effect is somewhat larger.

The effect of TMCs on the PDF uncertainty, illustrated in
Fig.~\ref{fig:PASOPE} for the OPE prescription, is to reduce the
uncertainty band at large $x$ for the lower $Q^2$ value, in analogy
with the effect seen in Fig.~\ref{fig:NP_test} for the $R_{np}$ ratio,
with strength moving from lower $x$ to higher $x$ by the $x \to \xi$
rescaling of the structure functions.  The slightly different effects
of TMCs on the various structure functions present in the asymmetry
render the uncertainty band on $A_{\rm PV}$ more asymmetric at
$Q^2=2$~GeV$^2$.  At the higher $Q^2=10$~GeV$^2$ value, the impact
of TMCs on the uncertainty band is reduced considerably, with the
two bands (corresponding to no TMCs and the OPE TMC prescription)
approximately coinciding for all $x$.

The conclusion from the combined results of Figs.~\ref{fig:PAS} and
\ref{fig:PASOPE} is that the effect of TMCs and particularly their
uncertainties can be minimized in the $A_{\rm PV}$ ratio by measuring
the asymmetry at values of $Q^2 \sim 10$~GeV$^2$ or higher; at lower
$Q^2$, although the TMC uncertainties are not large, some residual
corrections will need to be applied in the range
$0.4 \lesssim x \lesssim 0.9$, where the TMCs are $\approx 1\%$ or
higher.

\subsubsection{Deuteron asymmetry}
\label{ssec:PVDIS-d}

Unlike for a proton target, for PVDIS on an isoscalar deuterium nucleus
most of the dependence on PDFs cancels if one assumes that PDFs in the
proton and neutron are related by charge symmetry \cite{Bj78}.
In fact, in the valence quark region ($x \gtrsim 0.5$) where sea
quarks and gluons can be neglected, the deuteron asymmetry can be
written at leading order as \cite{Hobbs08,PDG10,Ans94}
\begin{equation}
A_{\rm PV}^d\
\approx\ -\left( {G_F Q^2 \over 2 \sqrt{2} \pi \alpha} \right)
{6 \over 5}
  \left( g_A^e (2 g_V^u - g_V^d)\ +\ Y_3\, g_V^e (2 g_A^u - g_A^d)
  \right),\ \ \ \ \ \ [x \gg 0],
\label{eq:APV_Bj}
\end{equation}
where $g_V^u = -1/2 + (4/3) \sin^2\theta_W$,
      $g_V^d =  1/2 - (2/3) \sin^2\theta_W$,
      $g_A^u =  1/2$, and $g_A^d = -1/2$.
(Note that the conventions for $g_{V,A}^e$ in Ref.~\cite{Hobbs08}
differ by a factor 2 compared with those used here, although the
asymmetry is of course independent of the convention.)
Consequently accurate measurement of deuteron PVDIS has been proposed
as a sensitive test of either the weak mixing angle $\sin^2\theta_W$
(deviations of which from its Standard Model value may signal the
presence of new physics), or more conventionally of charge symmetry
violation (CSV) in PDFs.

Nonzero values of $\delta u$ and $\delta d$ are predicted in
nonperturbative models of the nucleon to arise from quark mass
differences and electromagnetic effects (for a review see
Ref.~\cite{CSVrev}), and can also be generated from radiative QED
corrections in $Q^2$ evolution \cite{MRSTCSV,MRSTQED,GJRQED}.
Defining charge symmetry violating PDFs by
\begin{equation}
\label{eq:dudd}
\delta u = u^p - d^n,\ \ \ \ \ \ \delta d = d^p - u^n,
\end{equation}
the PVDIS asymmetry (\ref{eq:APV_Bj}) in the presence of CSV
is modified according to
\begin{eqnarray}
\label{eq:CSV_g}
(2 g_{V,A}^u - g_{V,A}^d)
&\to& (2 g_{V,A}^u - g_{V,A}^d) (1 + \Delta_{V,A}),
\end{eqnarray}
where the fractional CSV corrections are given by
\begin{eqnarray}
\label{eq:CSV_Delta}
\Delta_{V,A}
&=&
\left( -\frac{3}{10} + \frac{2 g_{V,A}^u + g_{V,A}^d}{2 (2 g_{V,A}^u - 
g_{V,A}^d)}
\right)
\left( \frac{\delta u - \delta d}{u + d}
\right).
\end{eqnarray}
These approximate expressions serve to illustrate explicitly the role
of CSV in the PVDIS asymmetry; in practice, however, the full deuteron
asymmetry can be computed including the effects of CSV at NLO, as well
as sea quarks and gluons.

\begin{figure}[t]
\rotatebox{-90}{\includegraphics[width=5.7cm]{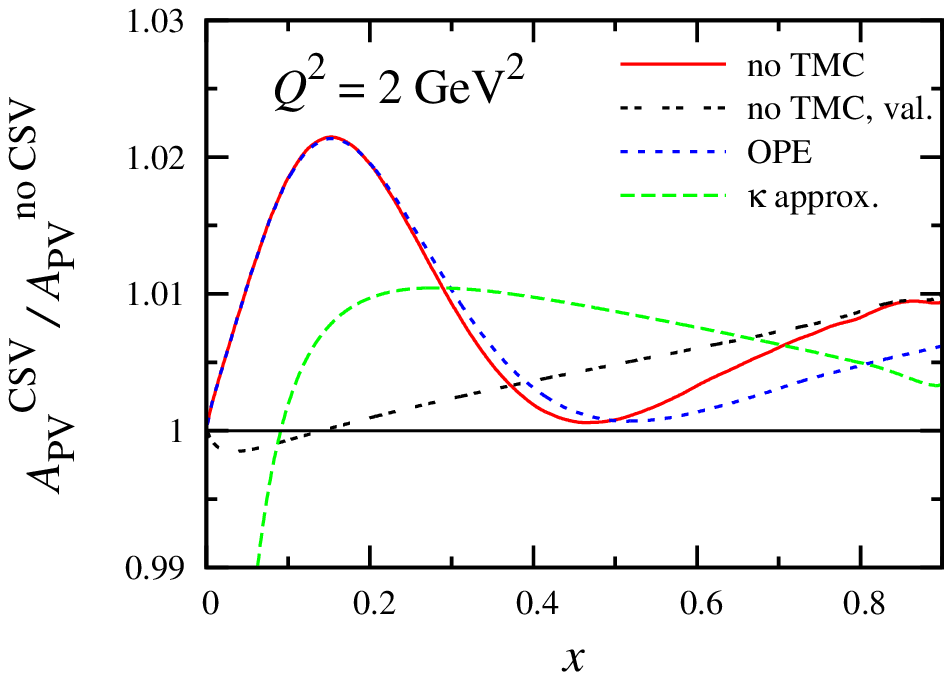}}
\rotatebox{-90}{\includegraphics[width=5.7cm]{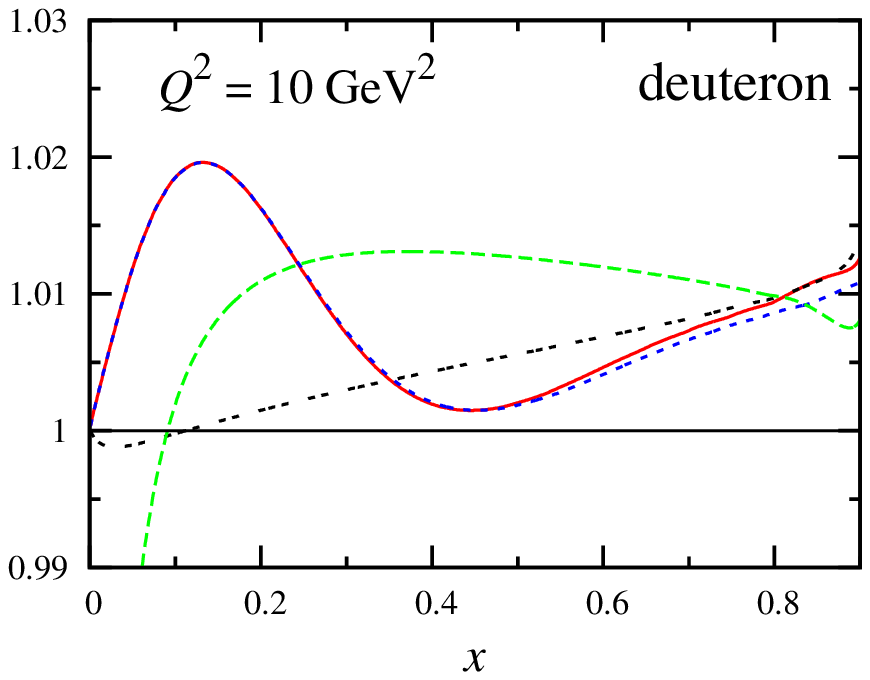}}
\caption{Deuteron PVDIS asymmetry including CSV effects, relative to the
	asymmetry with charge-symmetric PDFs, at $Q^2 = 2$~GeV$^2$
	{\bf (left)} and $Q^2 = 10$~GeV$^2$ {\bf (right)}.
	The CSV PDFs are computed from the MRSTQED parametrization
	\cite{MRSTQED} for the full asymmetry (solid, red) and for
	the valence approximation (double-dashed, black), and from
	the $\kappa$-dependent fit (see text) in Ref.~\cite{MRSTCSV}
	(long-dashed, green).  The effects of TMCs on the full asymmetry
	with the MRSTQED PDFs are illustrated for the OPE prescription
	(short-dashed, blue).}
\label{fig:CSV}
\end{figure}

Using the MRSTQED parametrization of PDFs \cite{MRSTQED}, which generates
nonzero values of $\delta u$ and $\delta d$ through radiative QED effects,
the effect of CSV on the deuteron asymmetry $A_{\rm PV}^d$ is illustrated
in Fig.~\ref{fig:CSV}.  In the valence quark region the CSV effect is
small at intermediate $x$, $x \sim 0.4$, but increases to around 1\%
at $x \sim 0.8$.  This is roughly comparable to the earlier fit in
Ref.~\cite{MRSTCSV} which parametrized the CSV PDFs as
$\delta u - \delta d = 2 \kappa \sqrt{x}(1-x)^4 (x-0.0909)$,
with $\kappa=-0.2$ as the best fit parameter.
(The constraints on $\kappa$ were found to be relatively weak,
however, and values of $\kappa$ from $-0.8$ to $+0.65$ produced fits
at the 90\% confidence level, with effects on the asymmetry ranging
from $\approx 4\%$ to 8\% over the range $0.4 \lesssim x \lesssim 0.8$.)
Deviations of the full NLO result from the valence approximation appear
already at $x \lesssim 0.7$, however, and these differ quite markedly
at small $x$, as Fig.~\ref{fig:CSV} indicates.  Interestingly, the full
asymmetry becomes larger at smaller $x$ because of CSV effects in the
light sea quarks, which produce an asymmetry of about 2\% at
$x \approx 0.2$.  On the other hand, cleanly separating the CSV effects
from sea quark and gluon contributions, which introduce additional $x$
dependence beyond that in Eqs.~(\ref{eq:APV_Bj}), (\ref{eq:CSV_g}) and
(\ref{eq:CSV_Delta}), as well as possible differences between CSV in
valence and sea quark PDFs, becomes more challenging at small $x$.

\begin{figure}[t]
\rotatebox{-90}{\includegraphics[width=5.7cm]{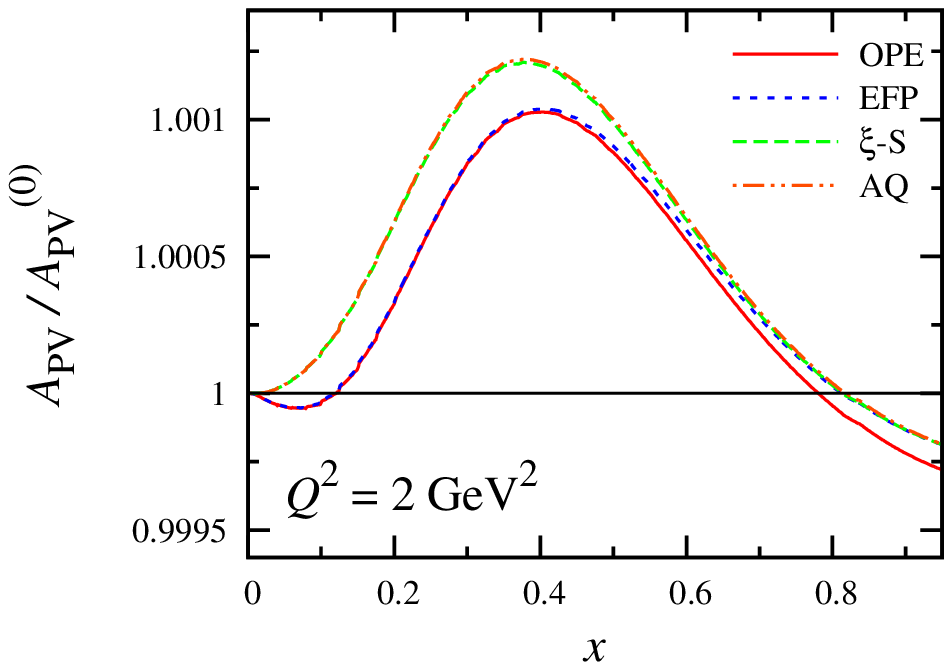}}
\rotatebox{-90}{\includegraphics[width=5.7cm]{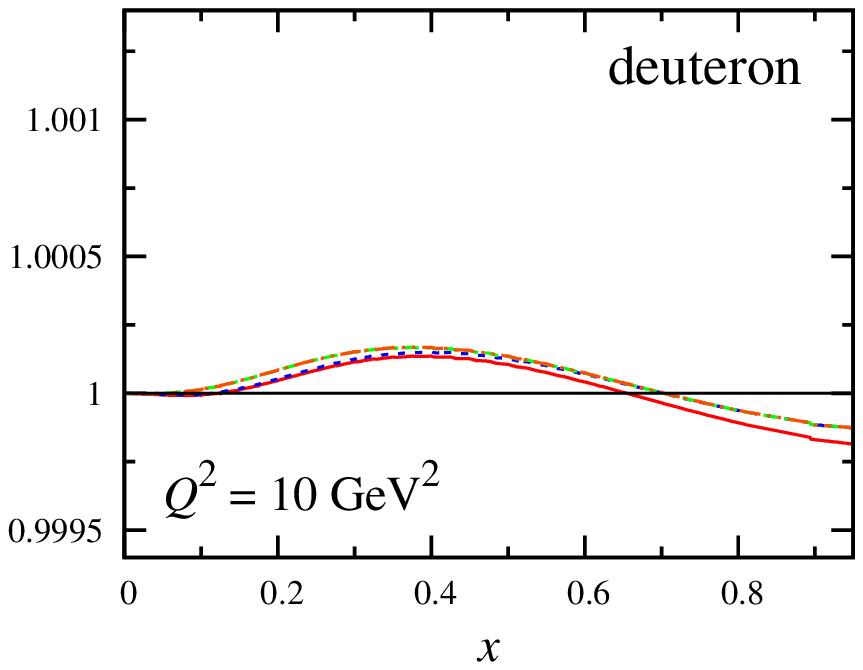}}
\caption{Ratio of target mass corrected to uncorrected PVDIS
	deuteron asymmetries $A_{\rm PV}^d$
	at $Q^2 = 2$~GeV$^2$ {\bf (left)} and
	$Q^2 = 10$~GeV$^2$ {\bf (right)}, for the
	OPE (solid, red),
	EFP (short-dashed, blue),
	$\xi$-S (long-dashed, green), and
	AQ (dot-dashed, orange) TMC prescriptions.}
\label{fig:DAS}
\end{figure}

With sought-after CSV effects that could be $\lesssim 1-2\%$,
it is vital to quantify the impact of TMCs on the deuteron PVDIS
asymmetries and in particular the TMC prescription dependence.
The effect of TMCs on the full asymmetry relative to the
charge-symmetric asymmetry is negligible at $x \lesssim 0.5$,
but decreases the CSV signal by up to 50\% at $x \approx 0.8$,
as Fig.~\ref{fig:CSV} demonstrates for the OPE prescription.
The model dependence of TMCs is illustrated for the various
prescriptions in Fig.~\ref{fig:DAS}, where the ratio of
asymmetries is shown with TMCs to those without TMCs.
The net effect is very small, peaking at $\sim 0.1\%$ at
$x \approx 0.4$, even at the $Q^2=2$~GeV$^2$ value.
The TMC prescription dependence of this ratio is even smaller, making
it essentially negligible on the scale of a CSV signal of $\sim 1\%$.
If the target mass corrected asymmetries were calculated with the
charge symmetry violating MRSTQED PDFs, the effect would be somewhat
larger, peaking at $\sim 0.3\%$ around $x \approx 0.4$.  However,
the TMC model dependence is still negligible at around 0.05\%.
As expected, the impact of TMCs on the deuteron asymmetries at the
larger $Q^2=10$~GeV$^2$ value is considerably smaller.  It is therefore
likely that TMCs would only play a role in deuteron PVDIS measurements
if the CSV effects were on the scale of a fraction of a percent,
at which point they would not be discernible within the expected
precision of the experiment \cite{SOLID}.  The corrections due to
nuclear smearing in the deuteron would have similarly negligible
effects on the measured deuteron asymmetry.

\section{Conclusions}
\label{sec:conc}

With the increased precision and kinematic reach of new experiments
planned in the next few years, particularly at Jefferson Lab with
its 12~GeV upgrade, the need for reliable theoretical tools with
which to analyze the data is becoming ever more pertinent.  This
is especially true for data that will be taken at large values of 
$x$, where a number of different subleading effects come to the fore.
In this work we have performed a comprehensive analysis of one class
of such corrections, namely those associated with finite values of
$x^2 M^2/Q^2$, or target mass corrections.
We have detailed several approaches to TMCs, including the standard
OPE method, as well as prescriptions based on collinear factorization,
and have compared their effects on various spin-averaged structure
functions at next-to-leading order.

For the TMCs computed via the OPE, we find that the $1/Q^2$ and $1/Q^4$
approximations to the full results are accurate only up to $x\approx 0.6$,
beyond which the series displays rather slow convergence.
Such an expansion has been proposed to avoid the threshold problem at
$x=1$; our findings suggest, however, that a low order expansion may
not be applicable as $x \to 1$.  Numerically, we find that TMCs in
the OPE approach are very similar to those computed via the EFP
implementation of collinear factorization, especially for the vector
structure functions $F_1$ (or $F_L$) and $F_2$.
This can be demonstrated analytically, through the equality
to order $1/Q^2$ of the prefactors associated with the leading terms.
The comparative phenomenology of these prescriptions has not previously
been addressed in the literature.

Similarly, the $\xi$-scaling and AQ prescriptions, which are derived
from different approximations within the collinear factorization
framework, yield corrected structure functions that closely track
each other over much of the $x$ range accessible experimentally.
In all cases the magnitude of the TMCs, and in particular their
model dependence is, not surprisingly, significantly more important
at low $Q^2$ values ($\sim 2$~GeV$^2$).  Target mass corrections
are suppressed with increasing $Q^2$, although even at
$Q^2 \sim 10$~GeV$^2$ they are not negligible for some observables
at very large $x$.  The greatest model dependence of TMCs arises
for the longitudinal structure function, where because of the
mixing between the $F_L$ and $F_2$ structure functions the effects
for the OPE and EFP prescriptions are significantly larger than
for the AQ and $\xi$-scaling approaches, where no mixing occurs.

In addition to quantifying the impact of TMCs on structure functions,
we further discussed the limitations these place on unambiguously
extracting information on PDFs (such as the $d/u$ ratio or charge
symmetry violation) from observables.  For the ratio $R_{np}$ of
neutron to proton $F_2$ structure functions we make the interesting
observation that at low $Q^2$ not only is one subject to greater TMC
uncertainties than at large $Q^2$, but the $x \to \xi$ rescaling due
to TMCs effectively also decreases the sensitivity to the $d/u$ ratio
at large $x$ that measurements of $R_{np}$ attempt to constrain.

For parity-violating DIS from the proton, the effects of TMCs and
perturbative NLO radiative corrections are similar in both the
electromagnetic and $\gamma Z$ interference LT ratios $R^\gamma$
and $R^{\gamma Z}$, with $\lesssim 4 - 5\%$ differences for
$Q^2 = 2$~GeV$^2$ at intermediate and larger $x$.  For the deuteron
the differences between $R^\gamma$ and $R^{\gamma Z}$ are smaller
in the valence quark dominated region, with negligible dependence
on the TMC prescription, but become larger at very small $x$
($\lesssim 8\%$ and 4\% at $Q^2 = 2$ and 10~GeV$^2$, respectively)
through gluonic contributions at NLO.  The magnitude of TMCs in the
$R^{\gamma Z}$ ratio itself, however, is significant at large $x$,
especially for the OPE and EFP prescriptions.  We also considered the
effects of nuclear corrections in the deuteron on the $\gamma Z$ LT
ratio, which become important for $x \gtrsim 0.4-0.5$; however, the
similarity of these with the effects on the electromagnetic LT ratio
leads to nuclear corrections largely canceling in the PVDIS asymmetry.

The effects of TMCs on the parity-violating asymmetries themselves
are generally rather small, especially at higher $Q^2$ values,
$Q^2 \sim 10$~GeV$^2$, although at lower $Q^2$ some residual
TMC dependence is evident in the case of the proton asymmetry.
Measurements of the proton PVDIS asymmetry are planned to provide
a unique combination of PDFs in order to constrain the $d/u$ ratio
at large $x$ \cite{SOLID}.  For the deuteron, the size of TMCs is
about an order of magnitude smaller than the expected CSV effects
in PDFs, which are estimated to be at the ${\cal O}(1\%)$ level.
On the other hand, while the corrections to the LT ratios and
asymmetries computed here have been perturbative, nonperturbative
effects such as those associated with nonzero parton transverse
momentum in the nucleon can produce additional strength in the
longitudinal structure functions \cite{Feyn72}.  This may be
particularly relevant for the ratio $R^{\gamma Z}$, whose phenomenology
is essentially unknown at low $Q^2$.  Estimates of nonperturbative
contributions to $R^{\gamma Z}$ would therefore be necessary before
making more definitive conclusions about its role in PVDIS.

In the future, additional effects not discussed here may need to be
considered at large $x$, principal among which are dynamical higher
twist corrections associated with nonperturbative multi-parton
correlations.  These are very difficult to compute from first
principles, and only rudimentary model estimates have been available
to date.  Further insight into the relation between TMCs and higher
twists may also shed light on the threshold problem, whereby the
target mass corrected structure functions remain finite at the nucleon
elastic scattering point, $x=1$, as well as on the difference between
the various TMC prescriptions.  Other corrections that may affect
future analysis of large-$x$ data are threshold resummations,
which involve formally summing, to all orders in $\alpha_s$,
terms containing logarithms of $1-x$ that become large as $x \to 1$.
The results on the phenomenology of the target mass corrections
contained in the present work should provide a benchmark for future
theoretical and experimental investigations of these additional
corrections.
This analysis can also be extended to the spin-dependent sector
\cite{Bluemlein99,Detmold06,AM08,Leader10}, where the phenomenology
of the collinear factorization framework in particular has not been
as fully developed.

\section*{Acknowledgements}

This work was supported by the DOE contract No. DE-AC05-06OR23177,
under which Jefferson Science Associates, LLC operates Jefferson Lab,
DoD's ASSURE Program, and the National Science Foundation under NSF
Contact Nos. 1062320 and 1002644.



\begin{thebibliography}{99}

\bibitem{MSTW08}
A.~D.~Martin, W.~J.~Stirling, R.~S.~Thorne and G.~Watt,
Eur.\ Phys.\ J.\ C {\bf 63}, 189 (2009).

\bibitem{JR09}
P.~Jimenez-Delgado and E.~Reya,
Phys.\ Rev.\ D {\bf 79}, 074023 (2009).

\bibitem{ABKM10}
S.~Alekhin, J.~Bl\"umlein, S.~Klein and S.~Moch,
Phys.\ Rev.\  D {\bf 81}, 014032 (2010).

\bibitem{HERAPDF10}
F.~D.~Aaron {\it et al.},
JHEP {\bf 1001}, 109 (2010).

\bibitem{CT10}
H.~-L.~Lai {\it et al.},
Phys.\ Rev.\  D {\bf 82}, 074024 (2010).

\bibitem{CTEQ6X}
A.~Accardi {\it et al.},
Phys. Rev. D {\bf 81}, 034016 (2010).

\bibitem{CJ11}
A.~Accardi {\it et al.},
Phys. Rev. D {\bf 84}, 014008 (2011).

\bibitem{NNPDF11}
R.~D.~Ball {\it et al.},
Nucl. Phys. {\bf B855}, 153 (2012).

\bibitem{SLAC}
L.~W.~Whitlow {\it et al.},
Phys.\ Lett.\ B{\bf 282}, 475 (1992).

\bibitem{Decade}
M.~E.~Christy and W.~Melnitchouk,
J. Phys. Conf. Ser. {\bf 299}, 012004 (2011).

\bibitem{BONUS12}
Jefferson Lab Experiment E12-10-102 [BONUS12],
S.~B\"ultmann, M.~E.~Christy, H.~Fenker, K.~Griffioen, C.~E.~Keppel,
S.~Kuhn and W.~Melnitchouk, spokespersons.

\bibitem{MARATHON}
Jefferson Lab Experiment E12-10-103 [MARATHON],
G.~G.~Petratos, J.~Gomez, R.~J.~Holt and R.~D.~Ransome,
spokespersons.

\bibitem{SOLID}
Jefferson Lab Experiment E12-10-007 [SoLID],
P.~Souder, spokesperson.

\bibitem{Brady11}
L.~T.~Brady, A.~Accardi, W.~Melnitchouk and J.~F.~Owens,
arXiv:1110.5398 [hep-ph].

\bibitem{Kuhlmann00}
S. Kuhlmann {\it et al.},
Phys. Lett. B {\bf 476}, 291 (2000).

\bibitem{Nachtmann73}
O.~Nachtmann,
Nucl.\ Phys.\  B {\bf 63}, 237 (1973).

\bibitem{GP76}
H.~Georgi and H.~D.~Politzer,
Phys.\ Rev.\  D {\bf 14}, 1829 (1976).

\bibitem{Schienbein08}
I.~Schienbein {\it et al.},
J.\ Phys.\ G {\bf 35}, 053101 (2008).

\bibitem{EFP83}
R.~K.~Ellis, W.~Furmanski and R.~Petronzio,
Nucl.\ Phys.\  B {\bf 212}, 29 (1983).

\bibitem{AOT94}
M.~A.~G.~Aivazis, F.~I.~Olness and W.~K.~Tung,
Phys.\ Rev.\  D {\bf 50}, 3085 (1994).

\bibitem{KR02}
S.~Kretzer and M.~H.~Reno,
Phys.\ Rev.\  D {\bf 66}, 113007 (2002).

\bibitem{AQ08}
A.~Accardi and J.~W.~Qiu,
JHEP {\bf 07}, 090 (2008).

\bibitem{Hobbs08}
T.~Hobbs and W.~Melnitchouk,
Phys.\ Rev.\ D {\bf 77}, 114023 (2008).
Note that in Eqs.~(9) and (13) the mass-dependent correction
factor should have the energy $E$ replaced by $2E$.

\bibitem{Hobbs11}
T.~Hobbs,
AIP Conf. Proc. {\bf 1369}, 51 (2011).

\bibitem{Bj78}
J.~D.~Bjorken,
Phys.\ Rev.\ D {\bf 18}, 3239 (1978).

\bibitem{Fajfer84}
S.~Fajfer and R.~J.~Oakes,
Phys.\ Rev.\  D {\bf 30}, 1585 (1984).

\bibitem{Castorina85}
P.~Castorina and P.~J.~Mulders,
Phys.\ Rev.\  D {\bf 31}, 2760 (1985).

\bibitem{Mantry10}
S.~Mantry, M.~J.~Ramsey-Musolf and G.~F.~Sacco,
Phys.\ Rev.\  C {\bf 82}, 065205 (2010).

\bibitem{Belitsky11}
A. V. Belitsky, A. Manashov and A. Sch\"afer,
Phys.\ Rev.\  D {\bf 84}, 014010 (2011).

\bibitem{PDG10}
K.~Nakamura {\it et al.},
J. Phys. G {\bf 37}, 075021 (2010).

\bibitem{Greenberg71}
O.~W.~Greenberg and D.~Bhaumik,
Phys.\ Rev.\  D {\bf 4}, 2048 (1971).

\bibitem{Tung79}
K.~Bitar, P.~W.~Johnson and W.~K.~Tung,
Phys.\ Lett.\  B {\bf 83} (1979) 114;
%
P.~W.~Johnson and W.~K.~Tung,
Print-79-1018 (Illinois Tech), Contribution to Neutrino '79,
Bergen, Norway (1979).

\bibitem{Steffens06}
F.~M.~Steffens and W.~Melnitchouk.
Phys.\ Rev.\ C {\bf 73}, 055202 (2006).

\bibitem{KP06}
S.~A.~Kulagin and R.~Petti,
Nucl.\ Phys.\  A {\bf 765} (2006) 126.

\bibitem{AHM09}
A.~Accardi, T.~Hobbs and W.~Melnitchouk,
JHEP {\bf 11}, 084 (2009).

\bibitem{Qiu90}
J.~-W.~Qiu,
Phys.\ Rev.\  {\bf D42}, 30-44 (1990).

\bibitem{Altarelli78}
G.~Altarelli and G.~Martinelli,
Phys.\ Lett.\  B {\bf 76}, 89 (1978).

\bibitem{Bardeen78}
W.~A.~Bardeen, A.~J.~Buras, D.~W.~Duke and T.~Muta,
Phys.\ Rev.\  D {\bf 18}, 3998 (1978).

\bibitem{CSS88}
J.~C.~Collins, D.~E.~Soper and G.~F.~Sterman,
Adv.\ Ser.\ Direct.\ High Energy Phys.\  {\bf 5}, 1 (1988).

\bibitem{MT96}
W.~Melnitchouk and A.~W.~Thomas,
Phys.\ Lett.\ B {\bf 377}, 11 (1996).

\bibitem{Afnan00}
I.~R.~Afnan {\em et al.},
Phys.\ Lett.\ B {\bf 493}, 36 (2000);
Phys.\ Rev.\ C {\bf 68}, 035201 (2003).

\bibitem{MEK05}
W.~Melnitchouk, R.~Ent and C.~E.~Keppel,
Phys. Rep. {\bf 406}, 127 (2005).

\bibitem{KMK09}
Y.~Kahn, W.~Melnitchouk and S.~A.~Kulagin,
Phys. Rev. C {\bf 79}, 035205 (2009).

\bibitem{EMC}
J.~J.~Aubert {\it et al.},
Phys.\ Lett.\ B {\bf 123}, 275 (1983);
%
D.~F.~Geesaman, K.~Saito and A.~W.~Thomas,
Ann.\ Rev.\ Nucl.\ Part.\ Sci.\  {\bf 45}, 337 (1995).

\bibitem{PVDIS-p}
P.~A.~Souder,
AIP Conf. Proc. 747, 199 (2005).

\bibitem{CSVrev}
J.~T.~Londergan and A.~W.~Thomas,
J. Phys. G {\bf 31}, 1151 (2005).

\bibitem{PVDIS-SLAC}
C.~Y.~Prescott {\it et al.},
Phys. Lett. {\bf 77} B, 347 (1978);
%
C.~Y.~Prescott {\it et al.},
Phys. Lett. {\bf 84} B, 524 (1979);
%
R.~N.~Cahn and F.~J.~Gilman,
Phys. Rev. D {\bf 17}, 1313 (1978).

\bibitem{Ans94}
M.~Anselmino, P.~Gambino and J.~Kalinowski,
Z.\ Phys.\  C {\bf 64}, 267 (1994).

\bibitem{MRSTCSV}
A.~D.~Martin, R.~G.~Roberts, W.~J.~Stirling and R.~S.~Thorne,
Eur.\ Phys.\ J.\  C {\bf 35}, 325 (2004).
        
\bibitem{MRSTQED}  
A.~D.~Martin, R.~G.~Roberts, W.~J.~Stirling and R.~S.~Thorne,
Eur.\ Phys.\ J.\  C {\bf 39}, 155 (2005).

\bibitem{GJRQED}
M.~Gluck, P.~Jimenez-Delgado and E.~Reya,
Phys.\ Rev.\ Lett.\  {\bf 95}, 022002 (2005);
arXiv:hep-ph/0501169.

\bibitem{Feyn72}
R.~P.~Feynman, {\it Photon-Hadron Interactions},
Benjamin, Reading, MA (1972).

\bibitem{Bluemlein99}
J.~Bl\"umlein and A.~Tkabladze,
Nucl.\ Phys.\ {\bf B553}, 427 (1999).

\bibitem{Detmold06}
W.~Detmold,
Phys.\ Lett.\  B {\bf 632}, 261 (2006).

\bibitem{AM08}
A.~Accardi and W.~Melnitchouk,
Phys.\ Lett.\ B {\bf 670}, 114 (2008).

\bibitem{Leader10}
U.~D'Alesio, E.~Leader and F.~Murgia,
Phys.\ Rev.\  D {\bf 81}, 036010 (2010).

\end{thebibliography}
\end{document}